\documentclass[useAMS,usenatbib]{mn2e}
\usepackage{times}
\usepackage{graphicx, amssymb}
    
\newcommand{\km}{${\rm km\,s}^{-1}$}
\newcommand\nodata{ ~$\cdots$~ }%

\newcommand{\hi}{H$\;${\small\rm I}\relax}
\newcommand{\hii}{H$\;${\small\rm II}\relax}

\newcommand{\nni}{N$\;${\small\rm I}\relax}

\newcommand{\ari}{Ar$\;${\small\rm I}\relax}

\newcommand{\ci}{C$\;${\small\rm I}\relax}
\newcommand{\cii}{C$\;${\small\rm II}\relax}

\newcommand{\civ}{C$\;${\small\rm IV}\relax}
\newcommand{\nii}{N$\;${\small\rm II}\relax}
\newcommand{\niii}{Ni$\;${\small\rm II}\relax}
\newcommand{\nv}{N$\;${\small\rm V}\relax}
\newcommand{\oi}{O$\;${\small\rm I}\relax}

\newcommand{\ovi}{O$\;${\small\rm VI}\relax}

\newcommand{\si}{S$\;${\small\rm I}\relax}
\newcommand{\sii}{S$\;${\small\rm II}\relax}
\newcommand{\siii}{Si$\;${\small\rm II}\relax}
\newcommand{\siiii}{Si$\;${\small\rm III}\relax}
\newcommand{\siiv}{Si$\;${\small\rm IV}\relax}

\newcommand{\mgii}{Mg$\;${\small\rm II}\relax}

\newcommand{\feii}{Fe$\;${\small\rm II}\relax}
\newcommand{\Niii}{Ni$\;${\small\rm II}\relax}
\newcommand{\alii}{Al$\;${\small\rm II}\relax}
\newcommand{\aliii}{Al$\;${\small\rm III}\relax}
\newcommand{\feiii}{Fe$\;${\small\rm III}\relax}

\title[Multiphase gas  in a $z\approx 2.4$ DLA]{Probing feedback in protogalaxies: Multiphase gas in a DLA at $z\approx 2.4$}

\author[N. Lehner et al.]{N.~Lehner,$^1$ J.C.~Howk,$^1$ J.X.~Prochaska,$^2$  and A.M.~Wolfe$^3$
\\
$^1$Department of Physics, University of Notre Dame, 
225 Nieuwland Science Hall, Notre Dame, IN, USA\\
$^2$UCO/Lick Observatory, University of California, Santa Cruz, CA, USA \\
$^3$ Department of Physics and Center for Astrophysics and Space Sciences, 
University of California, La Jolla, CA, USA
}
\date{Accepted XXX.
      Received XXX.}

\pagerange{\pageref{firstpage}--\pageref{lastpage}}
\pubyear{2006}

\begin{document}

\maketitle

\label{firstpage}

\begin{abstract}
We investigate the physical processes occuring in the multiphase gas of a damped Ly$\alpha$ system (DLA). 
We base our analysis on a high quality Keck HIRES spectrum  of the QSO J1211+0422 in which a DLA 
is detected at $z\simeq 2.377$. There is little contamination of the high-ion (\ovi, \nv, \civ, \siiv)
absorption, allowing us to  explore the properties of the highly ionized gas and its connection 
to other gas-phases.  The metallicity ($[{\rm Z/H}] = -1.41 \pm 0.08$), \hi\ column density 
($\log N($\hi$)= 20.80 \pm 0.10$), full-width velocity ($\Delta v_{\rm neut} \simeq 70$ \km) and relative
abundances ($[{\rm Si/Fe}] = +0.23 \pm 0.05$ and $[{\rm N/Si}] = -0.88 \pm 0.07$)  of this DLA are not unusual. 
However, we derive the lowest \cii*\ cooling rate in a DLA, $l_c < 10^{-27.8}$ erg\,s$^{-1}$ per H atom 
(3$\sigma$). Using this stringent limit, we show that the neutral gas (confined at $|v| < +39$ \km) must be warm and 
the star formation rate is $<7.1\times 10^{-3}$ M$_\odot$\,yr$^{-1}$\,kpc$^{-2}$. 
Surprisingly, the gas shows strong, complex absorption profiles from highly ionized gas
whose kinematics appear connected to each other and the low ions. 
The total amount of highly and weakly ionized gas is very large with 
$N_{\rm tot}($\hii$)/N_{\rm tot}($\hi$) \ga 1.5$. At $|v| \ga +39$ \km, 
the gas is fully and highly ionized (H$^+/$H\,$\sim 1$,  
$N($\civ$)\gg N($\cii$)$, $N($\siiv$)\gg N($\siii$)$). 
Based on ionization models, \ovi\ and \nv\ are generally difficult 
to produce by hard photons, while \siiv\ and \civ\ can be photoionized to a large extent. 
There is, however, no evidence of \ovi-bearing gas at $T\sim 10^6$~K
associated with this DLA. In contrast, there is some evidence for narrow \ovi, \nv, and \civ\  
components (unexplained by photoionization), implying too low temperatures ($T < 10^5$~K)  
for simple collisional ionization models to produce their observed 
column densities. Stellar feedback is a possible source for producing the high ions, but we cannot rule out 
accretion of non-pristine material onto the protogalaxy. 
\end{abstract}


\section{Introduction}
Damped Ly$\alpha$ absorbers (DLAs) are defined as absorbers with 
$\log N($\hi$) \ge 20.3$  \citep[][and references therein]{wolfe05}. 
As such these absorbers are often treated as dominantly neutral entities. 
This characteristic separates the DLAs from the other intervening
absorbers seen in the QSO sightlines, the so-called Ly$\alpha$ forest ($\log N($\hi$) \le 
17$) and Lyman limit systems  ($17 < \log N($\hi$) < 20.3$ ) where the neutral
gas is a minor or non-existent phase. The presence of neutral, cold, and molecular gas is crucial
to link the DLAs to star-forming galaxies  \citep{wolfe03,wolfe05,howk05,ledoux03,noterdaeme08}, 
although the exact nature of the DLAs and their relation to present-day galaxies is still under 
debate and study \citep{wolfe05,wolfe06,rauch08,meiksin08}. 

Surveys of DLAs have also shown systematic absorption from weakly 
(e.g., \aliii) and highly (\siiv, \civ, \nv, \ovi) ionized species in 
DLAs \citep{lu96,wolfe00a,fox07a,fox07}, demonstrating a complex interstellar
structure as observed in present-day galaxies. 
The detection of ionized gas, and more specifically highly ionized gas, 
is important because while high ions can be produced by hard photoionization, 
their ionization potentials are also high enough to be directly associated with galactic feedback 
following supernovae and stellar winds from massive stars \citep[e.g.,][]{kawata07}.

Stellar feedback is important since it may produce galactic outflows that are the dominant
mechanisms for enriching the intergalactic medium (IGM) with metals, as other 
dynamical processes, including ram-pressure stripping or tidal interactions between 
galaxies, are rather inefficient at polluting the IGM on large scales \citep[e.g.,][]{aguirre01}.
Energetic feedback from stellar winds and supernovae not only pollutes the IGM but is
also believed to serve as a regulator of star formation, even cutting it off altogether
by super heating or removing (almost) all the gas from a galaxy. Hence stellar 
feedback is an essential ingredient for cosmological models of the formation
and evolution of galaxies \citep[e.g.,][]{mashchenko08,dalla08}. 
For example, \citet{white91} showed that without 
feedback, star formation is far too efficient compared to the observations. However, 
feedback is extremely difficult to model because it requires enough resolution
to resolve individual supernova  and the ability to model the complex hydrodynamic 
processes in a highly multiphase interstellar gas \citep[e.g.][]{dalla08}. 
Characterizing feedback signatures via observations of the low and high $z$ Universe
is therefore critical. 

At all redshifts, evidence for outflows resulting from feedback processes 
has been discovered.  High-$z$ Lyman break galaxies (LBGs) show signs of 
outflows \citep{pettini01,shapley03}, analogous to the complex, 
multiphase outflows seen in low- and intermediate-$z$ starburst galaxies 
\citep{heckman00,martin06,tremonti07,weiner08}. At low redshift, 
winds have been found in even some modestly star forming galaxies such as 
the Large Magellanic Cloud \citep{lehner07} and the Milky Way 
\citep{zech07,yao07,bland03}.  

At $z\ga 2$, two
important DLA surveys undertaken by \citet{fox07a,fox07} have revealed the statistical
properties of the high ions associated with DLAs. They presented circumstantial 
evidence that \civ\ and \ovi\ absorption at high velocities relative to the 
DLAs is associated with protogalactic outflows. However, there is not yet a detailed analysis 
of the high ions (\ovi, \nv, \civ, and \siiv) in DLAs in order to test their origin(s) 
\citep[although for detailed analysis of a DLA with \civ\ and \siiv\ absorption, see][]{prochaska02}.
Part of the difficulty is that, while \civ\ 
and \siiv\ have generally little contamination, \nv\ absorption is weak
and \ovi\ is often mixed up in the Ly$\alpha$ forest. Among the 
9 QSOs with intervening DLAs presented by \citet{fox07}, there is not a single intervening DLA
where these four ions and both transitions of each doublet are simultaneously detected, 
complicating the physical interpretation of their profiles. In particular,
it is not clear if the apparent differences in breadth between the \ovi\ and \civ\  component-profiles 
are because \ovi\ probes truly hot gas or the
\ovi\ profiles are significantly blended with intervening Ly$\alpha$ absorbers. 

In this work, we analyze a Keck/HIRES spectrum of the QSO J1211+0422 
(SDSS\,J121117.59+042222.3, $z_{\rm em} = 2.5412$) in order to study 
the DLA  at  $z\simeq 2.377$ observed along this sightline. 
This DLA, not surprisingly, shows high-ion absorption. Remarkably,
however, the profiles of the high ions, as well as other tracers of neutral and weakly ionized 
gas, have little contamination from other intervening absorbers, 
allowing us to pursue detailed modeling of their profiles. 
Notably, both lines of the \nv\ doublet, while weak, are well detected. 
The \ovi\ doublet is partially tainted by Ly$\alpha$ forest, but there is enough 
information in each transition to interpret the \ovi. 
With our detailed modeling, the main issues we address
in this work are the connection  of the highly ionized with
the neutral and weakly ionized gas, the origin(s) of the highly ionized
gas in DLAs at  $z\sim 2.4$, and a search for possible feedback signatures. 

The outline of this paper is as follows: in \S\ref{sec-obs} we describe
the observation, give an overview of the velocity structure in the DLA,
present the analysis techniques and results (kinematics, column densities,
and $b$-values of the high ions). In 
\S\ref{sec-prop} we present and discuss the results for the 
low-velocity gas ($|v| < +39$ \km), while in \S\ref{sec-ion} 
the higher-velocity ($|v|\ge +39$ \km), fully ionized gas is presented. 
In \S\ref{sec-disc} we discuss some of the properties of this DLA 
in the context of other studies. Our main results are summarized in \S\ref{sec-sum}.

\section{Observations and Analysis}\label{sec-obs}
\subsection{Data reduction}

The data presented here were collected using the upgraded High
Resolution Echelle Spectrometer (HIRES; Vogt et al. 1994) on the Keck
I telescope on 3 May 2005.  The total exposure time was 2.5 hours.
The HIRES upgrade that took place in 2004 replaced the previous single
CCD with a new three CCD mosaic, including two CCDs with enhanced blue
sensitivity compared with the earlier instrument.  The
$2048\times4096$ MIT/Lincoln Labs CCDs (15 \micron\ pixels) were
binned by 2 pixels ($0\farcs24$) in the spatial direction and have 1.3
\km\ pixels in the dispersion direction.  All data were collected
using the UV cross disperser and the C1 decker, which has a
$0\farcs86$ slit width, giving a spectral resolution of $R\approx
45,000$ or $\approx$6.6 km/s for a source filling the slit.  The
seeing was approximately $0\farcs8$.

The reduction and extraction of the raw data used the HIRedux package
(v2.2) of J.X. Prochaska, which is distributed with the XIDL
routines.\footnote{Available through:
http://www.ucolick.org/$\sim$xavier/IDL/.}  We refer the reader to
the documentation for details.  In short, the two-dimensional echelle
images were bias-subtracted, flat-fielded, and wavelength-calibrated
using the HIRES ThAr and quartz (flat field) lamps.  One-dimensional
spectra are extracted using an optimal extraction routine, and
individual exposures and orders were co-added with an inverse variance
weighting.  A continuum was fit to each order before coaddition to
remove the effects of the blaze. The signal-to-noise per resolution 
element is between 35 and 45. For the figures showing the profiles
and apparent column densities, the spectra were binned by 3 pixels
(however, the profile fitting and apparent optical depth measurements 
described below were realized with the fully sampled spectra).

\subsection{Kinematics overview}
\begin{figure}
\includegraphics[width = 8.truecm]{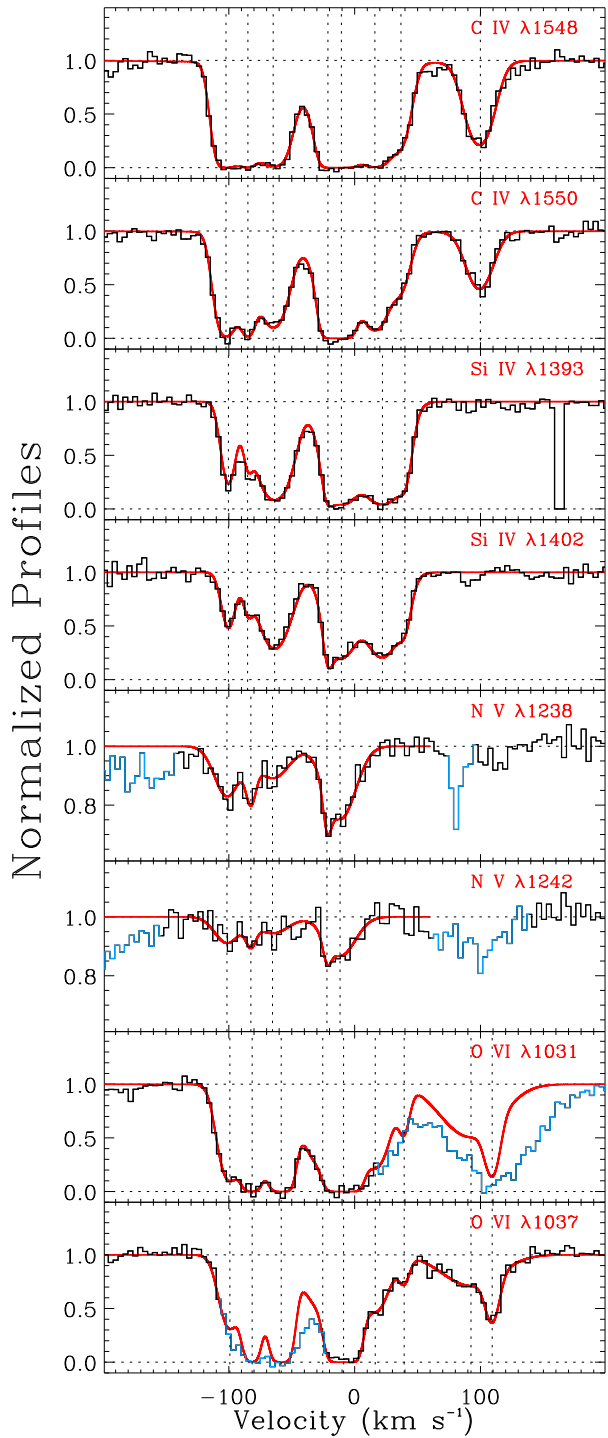}
\caption{Normalized profiles (histograms) of the highly ionized species observed 
at $z = 2.37656$ with their profile fits (smooth red curves). The vertical dotted 
lines show the centroid-velocities obtained from the profile fitting for each species.  
Each doublet was fitted simultaneously, but each species was fitted independently. 
Blue portions indicate some blending with unrelated lines. 
\label{spec1}}
\end{figure}

\begin{figure*}
\includegraphics[width = 15.5truecm]{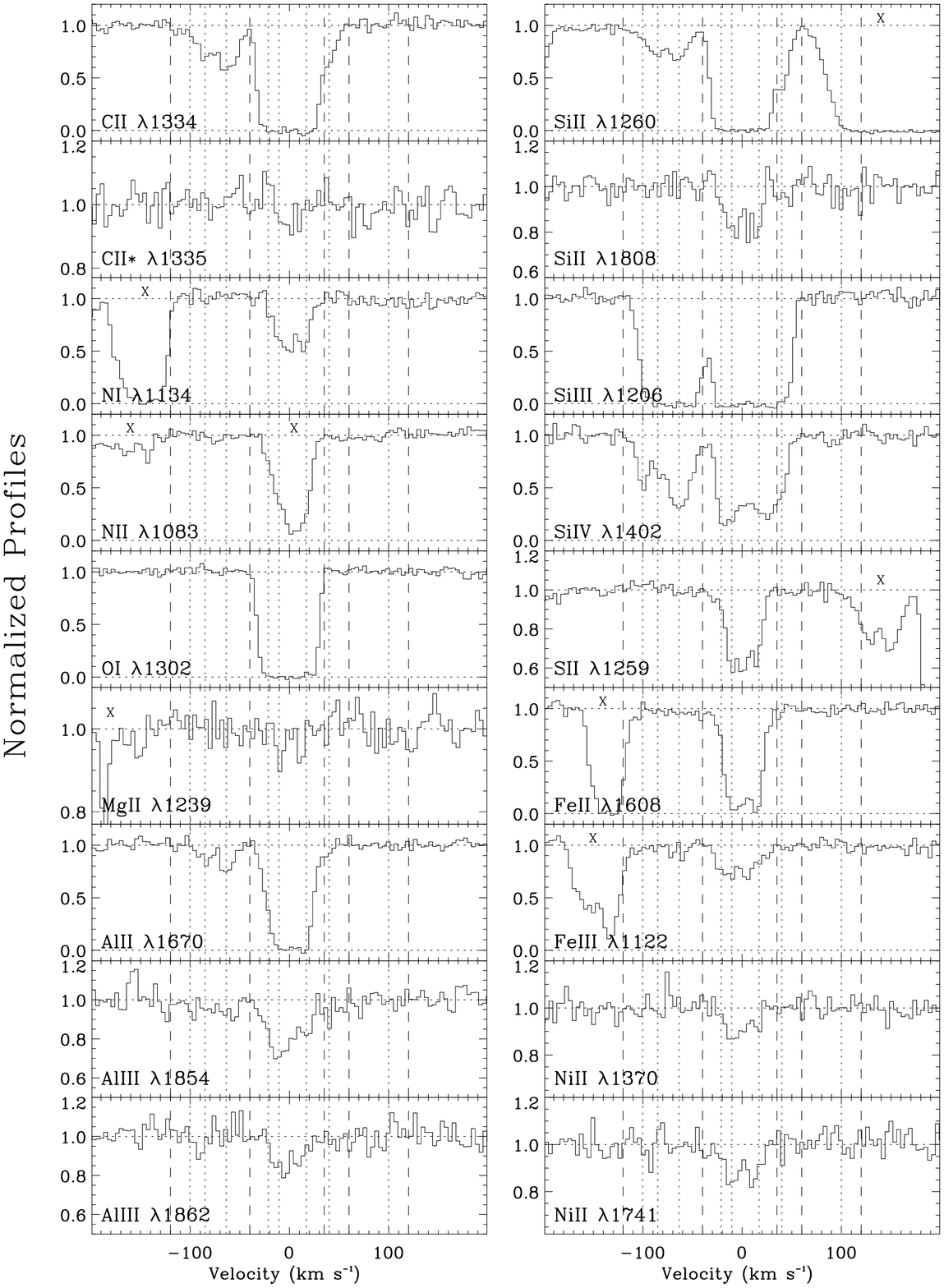}
\caption{Normalized profiles of several neutral and ionized species observed in the 
spectra of J1211+0422 at $z = 2.37656$. The vertical dashed lines show the separation
between the 4 main components: negative high-velocity component ($-120 \la v \la -40$ \km), 
main DLA component ($-40\la v \la +35$ \km), the $+39$ \km\ component,  
and positive high-velocity component only observed in \civ\ and \ovi\ ($+60 \la v \la +120$ \km, see Fig.~\ref{spec1}).  
The vertical dotted lines display the centroid-velocities of \siiv\ and \civ\ obtained from 
profile fitting. Unrelated lines are marked with crosses. Note that the $y$-scale changes from panel to panel.
\label{spec2}}
\end{figure*}

The normalized profiles of \civ, \siiv, \nv, \ovi\ are shown in Fig.~\ref{spec1}
against the rest-frame velocity at $z = 2.37656$ (determined from \siii\ $\lambda$1808 
and \oi\ $\lambda$$\lambda$1039,1302), while in Fig.~\ref{spec2}, 
we show the normalized profiles of the neutral and weakly ionized species. 
The profiles show complex kinematics that we separate into four main absorption
components according to their velocity spread: the negative high-velocity
component (NHVC) with $-120 \la v \la -40$ \km, the low velocity component (LVC) 
with $-40\la v \la +35$ \km,  the $+39$ \km\ component, and the 
positive high-velocity component (PHVC) with $+60 \la v \la +140$ \km. 

The LVC is associated with the main DLA component, which is in part neutral (strong saturated
\oi), but has also weakly and highly ionized species present. The NHVC is
more ionized than the LVC since no \oi\ or \nni\ absorption is detected and 
is also highly ionized since \civ\ absorption is saturated and the
strong transition of \cii\ $\lambda$1334 is not. For similar reasons, 
the $+39$ \km\ component also appears nearly fully and highly ionized. 
Finally, the PHVC is very highly ionized
with only detection of \civ\ and \ovi\ (and possibly \nv). In this work, we 
would like to understand the relationship (if any) of the observed highly ionized 
species with the other neutral and low ionized species (in particular, if relatively 
simple ionization models can reproduce the observed properties of the ionized gas), and
discern the origin(s) of these different components and the relationship with each other (if 
any). 
This requires that we first model the high-ion absorption, which we detail in the next section. 

\subsection{Kinematics and column densities of the high ions}\label{sec-fit}
To determine the basic properties of the high ions, our approach is twofold:
(1) a detailed modeling of the absorption assuming the gas can be modeled by Voigt 
profiles,
(2) a simpler approach using the apparent optical depth method.\footnote{To estimate the 
column densities in this work,
we use the atomic parameters compiled by Morton (2003), except \Niii\ $\lambda$1370 that 
is from  \citet{jenkins06}. } 

The absorption profiles of the high ions have several components, which 
we model using the Voigt component fitting software of 
Fitzpatrick \& Spitzer  (1997) where the instrumental line spread function was
modeled with a Gaussian with a full-width at half maximum of 6.6  \km.   
Each doublet was fitted simultaneously,
but each ionic species was fitted independently (except for two strong \civ\ components 
where we fix their velocities from \siiv, see below). We started by fitting the \siiv\ 
doublet because it is strong enough to see most of the components but not
so strong that saturation is a problem. Our  
strategy was first to determine the obvious components 
($\sim -100,-85,-65, -15,+20,+40$ \km)
and then add additional components, keeping them if the $\chi^2$ of the fit
improved significantly. This resulted in the first 7 components summarized
in Table~\ref{t1}. We note that the fit has some discrepancies in the core 
of the NHVC components in the \siiv\ $\lambda$1393  profile but
not observed in \siiv\ $\lambda$1402, which  we attribute to possible contamination 
from another intervening absorber (we therefore put more weight on \siiv\ 
$\lambda$1402 profile for the fit, but see below). There is also some 
discrepancy between the fit and the data near $-50$ and $-10$ \km, but adding 
more components did not improve the fit to the data. Yet since the main component 
of the DLA is near 0 \km, this could suggest the need to add an additional component 
near this velocity; 
doing so the fit produced a component at 3.6 \km, but with very uncertain
$b$-value and column density ($12.6 \pm 0.4$ dex) (note that adding
this component did not change significantly the parameters for the other components).
The reduced-$\chi^2$ also only changed marginally from 1.08 to 1.04. Thus, we adopt 
the result 
in Table~\ref{t1} as our best model. While the results of this fit may not be unique, 
they are robust since if we modify the input $b,N$ parameters, the solution of the 
fit converges to the one summarized in Table~\ref{t1}. However, it is important 
to bear in mind that the results from fitting complex profiles assume no hidden
components  (e.g., narrow or broad components) and the errors produced from the fitting code 
are only valid for the adopted number of components. That is a reason it is important to 
also investigate the profiles with another method (see below). 

The profiles of the \civ\ doublet are more complicated 
because the absorption is very strong in both lines. In particular, the flux from $-30$ to 
$0$ \km\ for both lines reaches zero, and therefore we 
force the centroid-velocities found from \siiv\ in this velocity range. 
The column densities and $b$-values in this velocity range 
remain, however, largely unknown.  Except for those 
two velocities, all the other parameters ($v,b,N$) were allowed to vary. 
The \civ\ profiles reveal an additional component not observed in \siiv, the so-called PHVC at
about $+100$ \km. We fitted  the PHVC with one component because while 
the over sampled profiles present some asymmetry, there is no real evidence for 
more than one component (see below). The overall fit to \civ\ is not as good as the model 
of \siiv\ because the \civ\ profiles are stronger. The fit reveals,
however, a very close correspondence in the kinematics between \civ\ and \siiv\ 
at least for the NHVC and the component at about 20 and 40 \km. 

The \nv\ doublet is much weaker but is clearly observed in both lines between
$\sim$$-110$  and  $\sim$$+20$ \km. Although there is little information 
on the $b$-values, the component-centroids are coincident with those 
of \siiv\ and \civ. The PHVC is partially contaminated in \nv\ $\lambda$1238
and fully contaminated in \nv\ $\lambda$1242. The column density reported in 
Table~\ref{t1} for the PHVC is estimated from the direct integration of the profiles 
(using the apparent optical depth method, see below) where the narrow and weak absorption at 
$+80$ \km\ was removed.

Finally, the \ovi\ $\lambda$$\lambda$1031, 1037 profiles are certainly the most 
complicated to model because they are so strong and because both lines are 
partially corrupted by other intervening absorbers. 
In regions where the profiles are contaminated (blue part of the spectra
in Fig.~\ref{spec1}), we use only one transition of the doublet 
and we assume that the used transition is not contaminated by intervening 
Ly$\alpha$ forest absorbers. The velocity-centroids match quite well the velocities of the other high ions
for $v<+50$ \km. At $v>0$ \km, \ovi\ $\lambda$1031 is blended with a strong absorber
and it is not clear if \ovi\ $\lambda$1037 could be partially blended as well 
since the velocity structure is different than that of \civ. We report in Table~\ref{t1}
a two-component fit to the PHVC, but owing to the uncertainty of possible blending, we 
will adopt the result from the apparent optical depth method (see below). However, 
it is worth noting that the $+109$ \km\ component is too narrow to be a Lyman intervening feature
\citep[see, e.g.,][]{kirkman97}.

\begin{table}
\begin{minipage}{8 truecm}
\caption{\large Summary of profile fitting measurements for the highly ionized species  
\label{t1}}
\begin{tabular}{lcccc}
\hline
Ion   & $v$  &   $b$ &   $\log N$	& Note \\
 & (\km)   &   (\km)  &   (cm$^{-2}$)    &
\\
\hline
\siiv\  & $-100.3 \pm 0.4 $  & $  6.1 \pm 0.7$   & $13.00 \pm 0.03$   & 	     \\
        & $ -85.1 \pm 0.6 $  & $  3.1 \pm 2.7$   & $12.55 \pm 0.07$   & 	     \\
        & $ -63.6 \pm 0.3 $  & $ 14.8 \pm 0.4$   & $13.56 \pm 0.01$   & 	     \\
        & $ -21.2 \pm 0.5 $  & $  2.8 \pm 1.2$   & $13.23 \pm 0.15$   & 	     \\
        & $ -10.6 \pm 0.7 $  & $ 14.1 \pm 0.6$   & $13.64 \pm 0.02$   & 	     \\
        & $ +22.1 \pm 0.7 $  & $ 15.5 \pm 1.1$   & $13.68 \pm 0.03$   & 	     \\
        & $ +39.9 \pm 0.6 $  & $  5.1 \pm 1.8$   & $12.77 \pm 0.13$   & 	     \\
        & $ +98.8 	  $  & \nodata	         & $<11.85        $   &  $a$	     \\
\hline
\civ\   & $-102.3 \pm 0.5 $  & $  8.2 \pm 0.4$   & $(>) 14.25 \pm 0.04$   & $b$	  \\
        & $ -84.8 \pm 0.5 $  & $  4.9 \pm 1.7$   & $(>) 14.16 \pm 0.17$   & $b$	  \\
        & $ -64.7 \pm 0.6 $  & $ 12.5 \pm 0.6$   & $(>) 14.13 \pm 0.02$   & $b$	  \\
        & $ -21.2	  $  & $  2.6 :$   & $(>) 17.27 :$  &  $b,c$	 \\
        & $ -10.6	  $  & $  7.7  \pm 3.8: $   & $(>) 14.86 \pm 0.53:$  &  $b,c$	 \\
        & $ +16.1 \pm 0.7 $  & $ 11.7 \pm 1.9$   & $14.16 \pm 0.04$   & 	  \\
        & $ +36.8 \pm 1.4 $  & $  8.6 \pm 1.2$   & $13.50 \pm 0.11$   & 	  \\
        & $ +99.1 \pm 0.2 $  & $ 12.9 \pm 0.5$   & $13.68 \pm 0.01$   & 	 \\
\hline
\nv\    & $-101.5 \pm 1.7 $  & $  13.0 \pm 3.4$  & $12.94 \pm 0.09$   &       \\
        & $ -82.5 \pm 1.1 $  & \nodata           & $12.48 \pm 0.41$   &      \\
        & $ -65.3 \pm 6.6 $  & \nodata    	 & $12.85 \pm 0.21$   &      \\
        & $ -22.0 \pm 0.8 $  & \nodata      	 & $12.42 \pm 0.30$   &      \\
        & $ -11.6 \pm 1.1 $  & $ 15.9 \pm 1.9$   & $13.21 \pm 0.05$   &      \\
        & $ +22.1  	  $  & \nodata  	 & $<12.55	  $   &  $d$	   \\
        & $ +39.9  	  $  & \nodata  	 & $<12.21	  $   &  $d$	\\
        & $ +98.5  	  $  & \nodata  	 & $(<)12.80:	  $   &   $e$  \\
\hline
\ovi\   & $-99.2 \pm 1.4 $   & $  9.5 \pm 1.1$   & $14.08 \pm 0.08$   & 	\\
        & $-81.5 \pm 0.9 $   & $  4.6 :	     $   & $(>) 15.28 :		$   & $b$	\\
        & $-58.4 \pm 0.6 $   & $  5.4 :	     $   & $(>) 15.67 : $   & 	$b$\\
        & $-25.4 \pm24.5 $   & $ 19.2 :      $   & $(>)14.13 :		$   & $b$	\\
        & $ -8.8 \pm 0.8 $   & $  6.8 :	     $   & $(>)16.22 :		$   & $b$	\\
        & $+16.4 \pm 4.7 $   & $ 13.2 \pm 6.9$   & $14.00 \pm 0.22$   & 	\\
        & $+39.4 \pm 1.0 $   & $  4.3 :      $   & $13.19 \pm 0.14$   & 	\\
        & $+92.5 \pm 2.3 $   & $ 29.5 \pm 2.7$   & $\le 13.99 \pm 0.04:$   &   $f$   \\
        & $+109.3 \pm 0.4 $   & $  6.0 \pm 1.1$   & $\le 13.74 \pm 0.05:$   &  $f$      \\
\hline
\end{tabular}
Note: The colons after a value or a 1$\sigma$ error indicate that the measurements are likely more 
uncertain than 
the formal errors derived by the profile fitting.
$a$: The 3$\sigma$ error is estimated using \siiv\ $\lambda$1393 based on the centroid 
and width of the \civ\ profile at this velocity. 
$b$:  The ``$(>)$'' indicates  that the absorption reaches zero flux in both lines and 
therefore those 
are likely lower limits and in all cases remain very uncertain.
$c$: For these components, we fix the velocity
centroids to those of \siiv.
$d$: The 3$\sigma$ error is estimated using \nv\ $\lambda$1238 based on the centroid 
and width of the \civ\ profile at this velocity (see \S~\ref{sec-fit} for more details). 
$e$: The "$(<)$" symbol indicates that \nv\ $\lambda$1238 is partially contaminated;
\nv\ $\lambda$1242 is completely contaminated and cannot be used. 
$f$: It is possible that \ovi\ $\lambda$1037 is partially contaminated by an intervening 
system, which is emphasized by the ``$\le$'' sign (see \S~\ref{sec-fit} for more details).
\end{minipage}
\end{table}

The $b$-value derived from the profile fitting can be used to 
estimate the temperatures of the high ions for each component. 
In Table~\ref{t2}, we list the temperature of the gas derived from the observed 
broadening of the absorption \siiv\ and/or \civ\ lines. If $b($\civ$)>b($\siiv) then $T$ can be 
determined from $b = \sqrt{2 kT/(Am_{\rm H}) + b^2_{\rm nt}}$, where $b_{\rm nt}$ is the non-thermal 
component and the other symbols have their usual meaning.
If $b($\civ$)<b($\siiv) only an upper limit on the temperature can be derived
because we cannot estimate the non-thermal broadening. We assume here 
that \civ\ and \siiv\ reside in the same gas (we will discuss in \S\S\ref{sec-prop}, \ref{sec-ion}
that this may not be always the case). We did not use \nv\ 
or \ovi, as their $b$-values are generally more uncertain and
these ions possibly exist in different phases (i.e. at different temperatures) than \civ\ or \siiv. 
Indeed the excitation potentials are 113.9 eV, 77.5 eV, 47.9 eV, and 33.5 eV
for \ovi, \nv, \civ, and \siiv, respectively. Therefore, \ovi\ and \nv\ are
a priori better diagnostics of shock-heated gas than \civ\ and \siiv, while photoionization 
is likely more important for \civ\ and \siiv. 
Half of the component sample may imply temperatures of a few times $10^5$ K gas 
if thermal broadening dominates. The other half implies temperatures much cooler, 
a few times $10^4$ K or less. 

\begin{table}
\begin{center}
\begin{minipage}{5.8 truecm}
\caption{Temperature$^a$ of the highly ionized \civ\ and \siiv\ gas \label{t2}}
\begin{tabular}{llccc}
\hline
Component & Name   & $T$ ($10^4$ K) \\
\hline
$-100$ \km\ & NHVC	  & $3.9 \pm 0.4$	       \\
$ -85$ \km\ & NHVC	  & $< 1.6$		       \\
$ -64$ \km\ & NHVC	  & $< 11$		       \\
$ -21$ \km\ & LVC	  & $< 1.3$		       \\
$ -11$ \km\ & LVC	  & $<34 $		       \\
$ +22$ \km\ & LVC	  & $< 10$		       \\
$ +39$ \km\ & $ +39$ \km\ & $<4.4  $		       \\
$+100$ \km\ & PHVC	  & $< 12$		       \\
\hline
\end{tabular}
Note: $a$: Temperature estimated from the broadening of 
\siiv\ and/or \civ\ ($b^2 = 0.129^2 \, T/A + b^2_{\rm nt}$, 
where $b$ and $b_{\rm nt}$ are in \km). 
\end{minipage}
\end{center}
\end{table}

We also explore the high-ions profiles with a method that does not rely on knowing the velocity
distribution in these profiles. The apparent optical depth (AOD) method 
(see Savage \& Sembach 1991) provides such a method. 
In this method, the absorption profiles are converted into
apparent column densities per unit velocity $N_a(v) = 3.768\times
10^{14} \ln[F_c/F_{\rm obs}(v)]/(f\lambda)$, where $F_c$ is the
continuum flux, $F_{\rm obs}(v)$ is the observed flux as a function of
velocity, $f$ is the oscillator strength of the absorption and
$\lambda$ is in \AA. This method also allows us to directly check
the saturation in the profiles by comparing the line of the same species
with different $f\lambda$, and in case the  absorption
is not saturated, $N_a = \int_{v_-}^{v_+} N_a(v) dv$ provides
a direct estimate of the true column density. Another advantage of this 
method is that it is straightforward to check for any variation in 
the ratio of $N_a(v)$ with velocities for various species. 

\begin{figure}
\includegraphics[width = 7.8truecm]{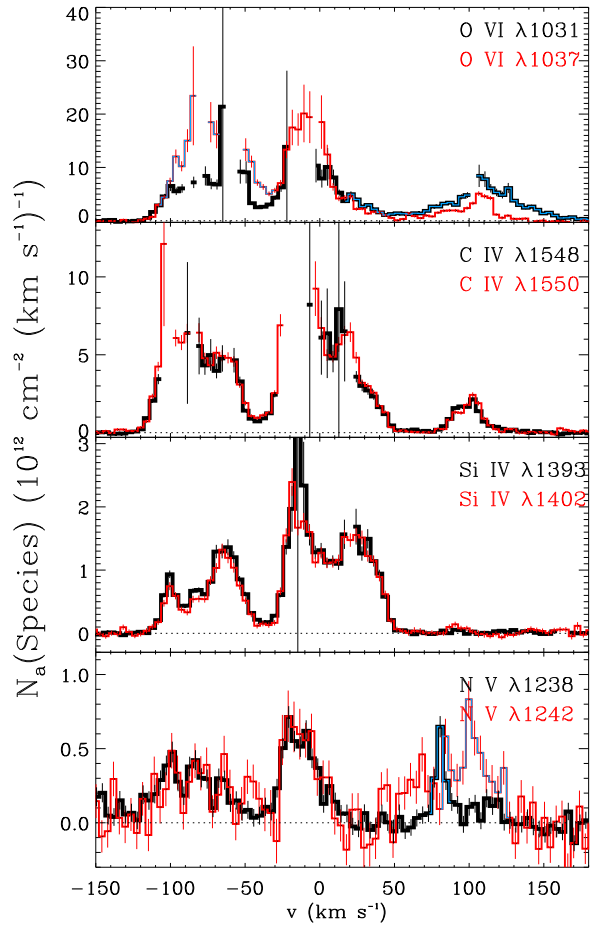}
\caption{Apparent column density profiles of the highly ionized species at $z = 2.37656$. Portions of 
profiles that appear blue show some contamination from another absorption feature.  
For velocities where the doublet AOD profiles match each other within the errors, 
the lines are resolved. Missing data occur when the flux reaches zero. 
\label{aodp}}
\end{figure}

In Fig.~\ref{aodp}, we show the apparent column density profiles for 
each doublet. There is a good overall agreement between the $N_a(v)$
profiles of \siiv\ $\lambda$1393 and $\lambda$1402, except in 
the cores of the various components, where \siiv\ $\lambda$1393 appears 
stronger; but this contamination is small since
since $N_a = \int_{-120}^{-40} N_a(v) dv = 10^{13.72 \pm 0.01}$ 
and $10^{13.66 \pm 0.02}$ cm$^{-2}$ and $N_a = \int_{-40}^{+60} N_a(v) dv = 
10^{14.06 \pm 0.03}$ 
and $10^{14.02 \pm 0.02}$ cm$^{-2}$ for \siiv\ $\lambda$1393 and $\lambda$1402, 
respectively. This shows that, despite the 
absorption being quite strong, the profiles of \siiv\ have essentially
no unresolved saturation. The \nv\ doublet is weak and fully resolved between
$-120$ and $+30$ \km. For $v >+40$ \km,  \nv\ $\lambda$1242 
is lost in another intervening absorber.  For \civ, 
only parts of the profiles are resolved. For \ovi, we further see 
that we need to rely on only one line in several portions of the absorption.

\begin{figure}
\includegraphics[width = 7.8truecm]{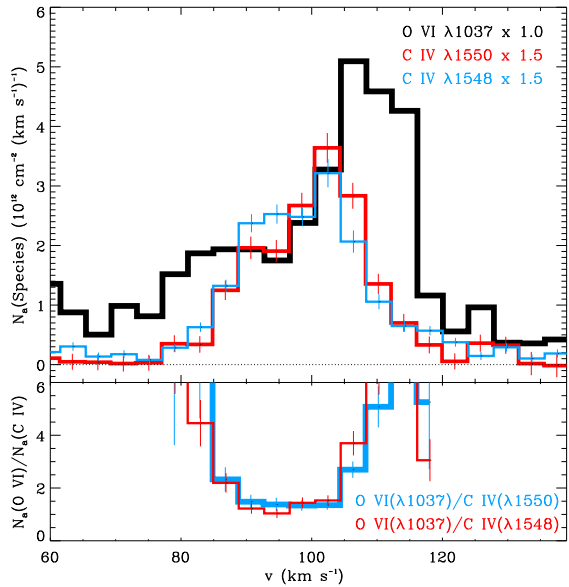}
\caption{Profile and ratios of the apparent column densities of \civ\ and \ovi\ 
for the positive high-velocity component with respect to $z = 2.37656$. At $v\la -85$ \km\ and $v\ga +105$ \km, 
\ovi\ might be contaminated. However, at $v\ga +105$ \km,  the feature
is too narrow to be a Lyman line. 
\label{aodhvc}}
\end{figure}

In Fig.~\ref{aodhvc}, we zoom in on the $N_a(v)$ profiles near the PHVC. 
The $N_a(v)$ profiles of the \civ\ doublet agree very well in the wing 
of the profiles. There is some discrepancy between $90$ and 107
\km, but not in a systematic manner, and  within the noise level 
both lines give a similar column density since 
$N_a = \int_{75}^{140} N_a(v) dv = 10^{13.68 \pm 0.01}$ and $10^{13.66 \pm 0.02}$
cm$^{-2}$ for \civ\ $\lambda$1548 and $\lambda$1550, respectively (note 
that the AOD result is fully consistent with the column density derived from the
one component fit).  This figure also shows clearly
the difference between \ovi\ and \civ\ with an increase in the \ovi/\civ\
ratio at the edge of the \civ\ profiles where the \ovi\ absorption is still
important. It is not clear if those differences 
at $+60\la v\la +85$ \km\ and $+105 \la v \la +120$ \km\ are real since 
\ovi\ $\lambda$1037 may be partially blended. 

\begin{figure}
\includegraphics[width = 7.8truecm]{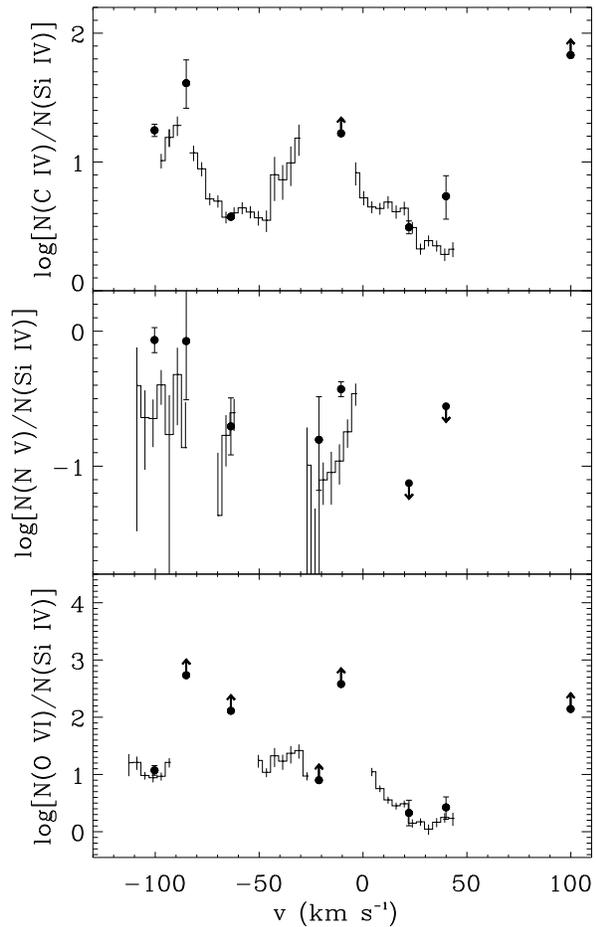}
\caption{Ratio of column densities at $z = 2.37656$. Filled circles are from the profile fitting method,
while histograms are from the apparent optical depth method. 
\label{aodr}}
\end{figure}

Finally, in Fig.~\ref{aodr}, we compare the column density ratios
of the high ions obtained from the AOD and profile fitting. We use
\siiv\  as the reference ion. For the AOD, we only plot part of 
the profiles that do not reach zero flux and we use weak lines
of the doublet for \civ\ and \siiv, the strong line of the \nv\ 
doublet, and non-contaminated part of the \ovi\ doublet. 
In view of the complexity of the profiles, there is 
a good overall agreement between the two methods (when the lines 
are not saturated), giving us confidence in the results of our fits
outside the regions where the profiles are completely saturated. 
However, we emphasize that the column densities reported in Table~\ref{t1} at $-21$ 
and $-9$ \km\ for \civ\ and \ovi, respectively, are totally uncertain. 
In the remaining of the text, we will only use the lower limits reported in 
Table~\ref{t4a} that are based on AOD estimates.

\section{The low-velocity multiphase gas}\label{sec-prop}

\subsection{Overview}
\begin{figure}
\includegraphics[width = 6.5 truecm, angle = 90 ]{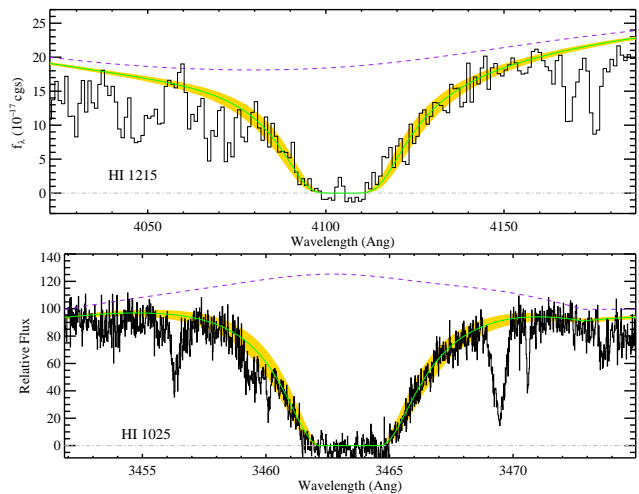}
\caption{Simultaneous fit to the damped Ly$\alpha$ (SDSS spectrum) and Ly$\beta$ 
(Keck HIRES) lines at $z = 2.37656$.  The dashed lines show the adopted continua. 
The bump in the continuum of Ly$\beta$ (bottom panel) is due to the blaze 
function of the echelle spectrum.  The solid green line show the solution 
$\log N($\hi$) = 20.80$ with the shaded region showing the $\pm 1 \sigma$ solutions ($\pm 0.10$ dex). 
Note that the wavelength scales are different in the top and bottom panels. 
\label{fithi}}
\end{figure}

We define the LVC as being the absorption observed 
between about $-40$ and $+35$ \km\ where most of the neutral 
gas is found. The  \oi\ profile in Fig.~\ref{spec2} traces the
velocity distribution of neutral gas, noting that
\oi\ is one of the best tracers of neutral hydrogen 
since its ionization potential and charge exchange reactions with hydrogen ensure that
\hi\ and \oi\ are strongly coupled. The LVC has 
a complex multiphase structure of neutral gas (\oi, \nni), 
ionized gas (\nii, \aliii, \feiii), and highly ionized gas (\ovi, \nv). 
The LVC differentiates itself from the other components at
higher absolute velocities by  the fact that it is only between 
$-40$ and $+35$ \km\ that strong atomic absorption (\oi, \nni) is observed;
in contrast the components at higher absolute velocities are nearly fully ionized as we 
show in \S\ref{sec-ion}. The presence of neutral gas allows us to directly
estimate the metallicity of the gas probed (see \S\ref{ssec-metal}).
In Table~\ref{t4}, we summarize our column density and abundance estimates for 
the LVC where for the species other than \hi, the column densities were estimated
via the AOD method  (see \S\ref{sec-fit}). The \hi\ column density is derived from 
a simultaneous fit to  the damped wings of Ly$\alpha$ (from SDSS) and Ly$\beta$
\citep[following][]{herbert06}, and we 
show the fit to those lines in Fig.~\ref{fithi}. 

For the high ions, the LVC has at least 3 components derived from the fits.  
Inspection to the \sii\ profile also reveals that the main absorption can be broken-up 
in 3 components with velocities centered at $\sim -10, 0, +15$
\km. Such structure cannot be seen in \oi\ because the profile is completely saturated, 
but the velocity structure of \nni\ is consistent with \sii. 
\oi\ reveals another  component at $+22$ \km, which is also seen in the high ions. 
In Fig.~\ref{spec2}, we show that except for the component at 0 \km\, the velocity-
centroids of the high ions match quite well those of the low ions, suggesting some kinematic 
relation between the weakly and highly ionized phases. The zero velocity component is not included in our
fits to the high ions, but as we discussed in \S\ref{sec-fit}, the fit for the 
high ions may not be unique and can certainly accommodate a component at $\sim 0$ \km. 

\subsection{Absolute and relative abundances}\label{ssec-metal}
To derive the metallicity, it is usually best to use neutral or singly-ionized species that
follow the neutral hydrogen to avoid large ionization correction since only $N($\hi$)$ can be directly estimated. 
The \oi\ absorption is so strong that only a firm lower limit can be placed on the oxygen abundance 
of the DLA (see Table~\ref{t4}). Because of the nucleosynthetic evolution, nitrogen is often deficient with respect 
to the other elements \citep[][and see below]{lu98,prochaska02a,henry00,henry06,petitjean08}, and this is observed in 
this DLA since  $[$N$/\alpha] = -0.88 \pm 0.07$, where the $\alpha$-element is 
here Si (see Table~\ref{t4} and also \S\ref{ssec-ion}). Nitrogen therefore does not provide 
a good metallicity indicator.

In the next section, we show that the singly-ionized 
species (excluding \nii) predominantly arise in the neutral gas, and 
therefore these ions can be directly compared to \hi. Sulfur appears not to be 
depleted into dust and silicon is usually very mildly depleted in DLAs, and $[{\rm Si/S}] = +0.02 \pm 0.06$
in this DLA suggests as well no dust depletion of Si. Using these
two elements, we derive the metallicity of the DLA, $[{\rm Z/H}] = -1.41 \pm 0.08$,
based on the $\alpha$-elements S and Si.

\begin{table}
\begin{minipage}{8.2 truecm}
\caption{\large DLA (LVC) column densities and abundances  \label{t4}}
\begin{tabular}{lcc}
\hline
Species X   & $\log N($X$)$ &  $[$X$/$H$]^a$   \\
\hline
\hi\ $\lambda$$\lambda$1025,1215	& $20.80 \pm 0.10$	& \nodata	   \\
\ci\ $\lambda$1560			& $< 12.70	 $	& ($<-4.49$)  	    \\
\cii\ $\lambda$1036			& $> 14.70	 $	& $>-2.49 	 $  \\
\cii*\ $\lambda$1335	 		& $< 12.53	 $	& ($<-4.66$)  	    \\
\civ\ $\lambda$1550			& $> 14.50	 $	& $(>-2.69) 	 $  \\
\nni\ $\lambda$1134			& $14.30 \pm 0.05$	& $-2.28 \pm 0.11$  \\
\nii\ $\lambda$1083			& $\le14.37 \pm 0.10$	& $(<)-2.21 \pm 0.15$  \\
\oi\ $\lambda$1039			& $> 15.32	 $	& $>-2.14 	 $  \\
\mgii\ $\lambda$1239			& $14.87\,^{+0.12}_{-0.17}$& $-1.46\,^{+0.19}_{-0.21} 	 $  \\
\alii\ $\lambda$1670			& $> 13.42	 $	& $>-1.80	 $  \\
\aliii\ $\lambda$$\lambda$1854,1862 	&$12.69 \pm 0.03$	& ($-2.56 \pm 0.11$) \\
\siii\ $\lambda$1808			& $14.91 \pm 0.04$	& $-1.40 \pm 0.11$  \\
\siiii\ $\lambda$1206			& $> 13.21	$	& ($>-3.10$)  	    \\
\siiv\ $\lambda$1402			& $13.96 \pm 0.04$	& ($-2.30 \pm 0.10$)  	    \\
\si\ $\lambda$1425			& $< 12.38	$	& ($<-3.57$)  	    \\
\sii\ $\lambda$$\lambda$1250,1259	& $14.53 \pm 0.04$	& $-1.42 \pm 0.11$  \\
\ari\ $\lambda$1066			& $< 13.10	$	&  $<-1.88$  	    \\
\feii\ $\lambda$$\lambda$1063,1144,1608	& $14.62 \pm 0.03$      & $-1.58 \pm 0.10$  \\
\feiii\ $\lambda$1122			& $13.97 \pm 0.03$	& ($-2.28 \pm 0.11$)  	    \\
\Niii\ $\lambda$$\lambda$1370,1709,1741	& $13.36 \pm 0.08$	& $-1.64 \pm 0.13$  \\
\hline
\end{tabular}
Note: Errors are 1$\sigma$. Upper limits (``$<$'') are 3$\sigma$ estimated over $[-30,+30]$ \km, full
velocity-width of the weak absorption (e.g., \siii\ $\lambda$1808). $a$: Throughout the 
text we use 
the following notation ${\rm [X/H]} = \log N({\rm X}^i)/N({\rm H}^0) - \log({\rm 
X/H})_\odot$ (solar abundances are from Asplund et al. 2006); if ${\rm X}^i$ is
a dominant ionization stage in neutral gas, $[$X$/$H$]$ indicates the abundance of a given species X. 
Estimates within parentheses do not represent a species in a dominant ionization stage, 
and therefore not the overall abundance of the gas (see text for more
details).  ``$\le$'' indicates that \nii\ is likely contaminated.
\end{minipage}
\end{table}

Nucleosynthesis and dust depletion both play a role in the gas-phase abundances of the elements in 
DLAs \citep[e.g.,][]{lu96,pettini99a,prochaska02}, but disentangling these effects is
extremely difficult and requires more elements than available here. However, 
the low  $[$N$/\alpha]$ ratio implies that nucleosynthesis is important
since N is not depleted into dust and \nni\ is not affected by ionization in this case. 
This low  $[$N$/\alpha]$ ratio is often explained as a time delay 
between the production and injection of N into the gas compared with $\alpha$-elements. 
The $\alpha$-elements (including O, Si, S) are mainly produced by
short-lived massive stars ($M>10M_\odot$) through explosions of Type II supernovae (SNe), 
while N in low metallicity gas is believed to be principally formed in intermediate-mass stars 
($3 M_\odot<M< 8 M_\odot$) and injected into the interstellar gas through stellar mass loss. The same 
low and intermediate-mass stars also produce the bulk of the Fe-group elements through
Type Ia SNe. Therefore the $\alpha$-element to Fe-peak element ratio should be
systematically enhanced in low metallicity environments. This is observed 
in low metallicity stars in our Galaxy, where  $[{\rm \alpha/Fe}] \approx +0.2$ \citep[e.g.][]{nissen07}. 
In the present DLA, $[{\rm \alpha/Fe}] = +0.23 \pm 0.05$ and $[{\rm \alpha/Ni}] = +0.24 \pm 0.09$ (where
$\alpha=$\,\siii, but as noted above the use of \sii\ would not have changed
the results). These values are consistent with others  in DLAs, 
although on the low side. The $[{\rm \alpha/Fe}]$ enhancement coupled 
with  $[{\rm Si/S}]\approx 0$ and   $[$N$/\alpha] \approx -0.9$ is consistent
with a Type II SN enrichment \citep{prochaska02,henry07}, i.e. the relative
abundances in this DLA could be explained by nucleosynthesis alone.

\begin{figure}
\includegraphics[width = 9truecm]{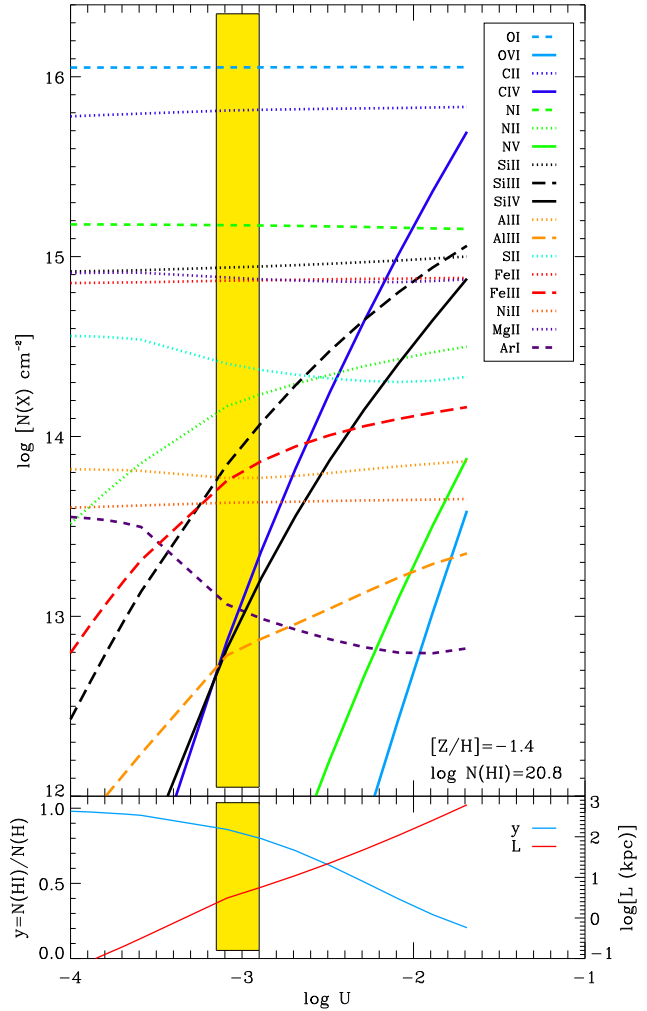}
\caption{{\em Top panel}: Predicted column densities for the Cloudy photoionization model of
the LVC  assuming a Haardt-Madau (galaxies+QSOs) spectrum. The various lines show the models for 
each atoms or ions. Relative solar abundances are assumed. The yellow region shows a solution
that fits the observations within about $\pm 0.2$ dex (see also Table~\ref{t4a}). This model does, however, 
not produce enough highly ionized species.  
{\em Bottom panel}: Variation of the neutral fraction and the path length ($L \equiv N({\rm H})/n_{\rm H}$).
For the solution that fits the observations, the fraction of neutral gas is 80--88\% and the linear scale is 
2--6 kpc.
\label{cloudy-lvc}}
\end{figure}

On the other hand, in our Galaxy, elements become less and less depleted as the gas changes 
from cold to warm phases \citep[see, e.g.,][]{savage96}. 
Assuming now that nucleosynthetic effects are negligible,
the low depletion of Fe and Ni with respect  to those in other DLAs {\em with similar metallicities} 
suggests that the neutral gas is likely to be mostly warm. As we show 
below the Cloudy simulation and the absence of \cii*\ absorption also support that the 
neutral gas is warm. An intrinsic solar $[{\rm \alpha/Fe}]$ at $[{\rm Z/H}] \approx -1.4$ 
may imply that the star formation occured slowly, allowing a large Fe pollution from
type Ia SNe \citep{matteucci01}, but this contradicts the extremely low $[{\rm N/\alpha}]$
ratio. More likely, the regions where star formation occurs are very localized and are not directly probed by 
the J1211+0422 sightline. Hence the gas may be indeed warm with little dust along the line of sight. 
Both effects from nucleosynthesis and dust depletion may be present, but it is not possible
to disentangle these subtle effects at the 20\% (0.1 dex) level (without mentioning possible
uncertainties in the atomic parameters or ionization
correction, see  \S\ref{ssec-ion}). The recent non-LTE abundances of S and Fe of
Galactic halo stars with $[{\rm Z/H}] \approx -1.4$ seem to imply that  $0.1 <[{\rm \alpha/Fe}] <0.2$
\citep{nissen07}, allowing both some Fe and Ni depletion and nucleosynthesis effects to occur
if the DLA nucleosynthetic history follows that of Galactic halo stars. In \S\ref{sec-cool}, 
we assume that the Fe depletion is $[{\rm Fe/Si}] = -0.2$ dex \citep[see discussion in][]{wolfe03,wolfe04}, which still
allows for an $\alpha$-element enhancement over Fe of $\sim$0.1 dex 
attributable to nucleosynthesis at the 1$\sigma$ level.

\begin{table}
\begin{minipage}{8.5 truecm}
\begin{center}
\caption{Column densities for the LVC from the Cloudy simulation
using solar relative abundances and $[{\rm Z/H}] = -1.4$ (Fig.~\ref{cloudy-lvc}).  \label{t4a}}
\begin{tabular}{lccc}
\hline
Species	& $\log N_{\rm obs}$ 	&   $\log N_{\rm model}$ &   $\log N_{\rm model}$      \\
	&			&   $\log U = -3.1$	 &   $\log U = -2.9$	       \\
\hline
\ci\		& $< 12.70	$	&	11.9     &    11.7     \\
\cii\		& $> 14.70	$	& 	15.8     &    15.8        \\
\cii*\ 		& $< 12.53	$	&	12.5     &    12.4    \\
\civ\ 		& $> 14.50	$	& 	12.9     &    13.4        \\
\nni\		& $14.30 \pm 0.05$	& 	15.2     &    15.2      	  \\
\nii\		& $<14.37 	$	&	14.2     &    14.2      	  \\
\nv\ 		& $13.28 \pm 0.05 $	&	10.6     &    11.2        \\
\oi\ 	 	& $> 15.32	 $	&	16.1     &    16.1         \\
\ovi\ 		& $> 14.86	 $	&	8.8      &    9.7         \\
\mgii\		& $14.87\,^{+0.12}_{-0.17}$& 	14.9     &    14.9         \\
\alii\ 		& $> 13.42	 $	& 	13.8     &    13.8         \\
\aliii\ 	&$12.69 \pm 0.03$	& 	12.8     &    12.9       \\
\siii\ 		& $14.91 \pm 0.04$	& 	14.9     &    14.9        \\
\siiii\ 	& $> 13.21	$	& 	13.8     &    14.1         \\
\siiv\ 		& $13.96 \pm 0.04$	& 	12.8     &    13.2    \\
\sii\ 		& $14.53 \pm 0.04$	& 	14.4     &    14.4         \\
\ari\ 		& $< 13.10	$	&  	13.1     &    13.0    \\
\feii\ 		& $14.67 \pm 0.03$      &	14.9     &    14.9        \\
\feiii\ 	& $13.97 \pm 0.03$	&	13.8     &    13.9    \\
\Niii\ 		& $13.36 \pm 0.08$	& 	13.6     &    13.6         \\
\hline
\end{tabular}
\end{center}
\end{minipage}
\end{table}

\subsection{Ionization}\label{ssec-ion}
The comparison of the apparent column densities of \siiv\ and \siii\ shows that the gas traced by
singly ionized species is far more important between about $-15$ and $+20$ \km.
In contrast the ratios of doubly-ionized
species (\aliii\ and \feiii) to \siiv\ show only little variations between 
$-40$ to $+40$ \km, suggesting some connection between these ions and that ionization 
may be important. The \nv/\siiv\ ratio  also steadily increases
from $+20$ to  $-40$ \km, implying that the gas becomes more highly ionized at more 
negative velocities. Although the \ovi\ absorption
is too strong to directly use for the column density ratios, the \ovi\ absorption is 
also far stronger at $-40 \la v \la 0$ \km\ than at $0 \la v \la +30$ \km\ 
(see Figs.~\ref{spec1} and \ref{aodp}).  
 
We first investigate whether  photoionization can explain the observed properties 
of the absorbers. We used the photoionization code 
Cloudy version C07.02 \citep{ferland98} 
with the standard assumptions  there has been enough time 
for thermal and ionization balance to prevail. 
We model the column densities of the different ions through a slab illuminated (on both 
sides) by the Haardt \& Madau (2008, in prep.) UV background  ionizing radiation field
from quasars {\em and}\ galaxies and the cosmic background radiation appropriate 
for the redshift $z = 2.377$ and maintained in this case at constant pressure 
(condition not set for the Cloudy simulations of the higher-velocity components).
The Cloudy simulations also assume a priori solar relative heavy element abundances 
from \citet{asplund06}. 
We then vary the ionization parameter, $U = n_\gamma/n_{\rm H} =$\,H ionizing photon density/total
hydrogen number density [neutral\,+\,ionized], to search for models that 
are consistent with the constraints set by the column densities and $b$-values. 

We show in Fig.~\ref{cloudy-lvc} the result of such a Cloudy simulation 
for the LVC where   $[{\rm Z/H}] = -1.4$ 
(see \S\ref{ssec-metal}) and  where the \hi\ column density matches
the measured value. We use the total column densities rather than modeling 
each component because i) the ratios of \feiii/\siiv\ and \aliii/\siiv\ show
little variation over the LVC velocity range, ii)
the total column densities are likely less uncertain than the column
densities of the individual components since they do not rely on 
a particular kinematic model, and iii) the \hi\ column density in the 
individual components is unknown.  The yellow region shows the range of possible
solutions  ($-3.15 \le \log U \le -2.90$) that fits the column densities of
the neutral, singly- and doubly-ionized species within $\pm 0.1$--0.2 dex. 
For $\log U = -3.1$ and $\log U = -2.9$, column densities for the various species 
are reported in Table~\ref{t4a}, showing an overall agreement between the model
and the observations, except for \nni. The low abundance of N  can be understood in terms
of nucleosynthetic evolution (see above): using \nni, we find $[$N$/\alpha] \simeq -0.9$
(\nii\ is contaminated by an unknown amount and is anyway important only in ionized gas). 
From Table~\ref{t4a}, we also note 
some discrepancies with \feii\ and \niii\ (although it is within $\sim 0.2$ dex), 
which can be interpreted as these species being in part depleted into dust or as a
nucleosynthetic effect (see \S\ref{ssec-metal}). However, if in the ionized phase Fe is depleted into
dust or deficient relative to $\alpha$-elements as a result of nucleosynthesis, 
the discrepancy between the model and observations would increase for \feiii, possibly
pointing to the need of another ionizing source (see below). 

The overall good agreement between the model and the observation is quite 
remarkable in view of the complexity of the profiles and the possible
effects of dust and nucleosynthesis. 
We note that although the agreement would be better for \sii\ if $\log U < -3.1$, 
smaller ionization parameters are inadequate because too much \ari\ would be produced. 
$[$\ari$/\alpha] < -0.48$ (where here $\alpha$ is \siii) favors the present model
rather than a softer stellar spectrum. Indeed, since Ar is not depleted into dust and 
is an $\alpha$-element, the observed \ari\ deficiency is due to photoionization \citep{sofia98,prochaska02}. 
In our own Galaxy, partially photoionized gas leads to a deficiency of 
the neutral form of Ar \citep{sofia98,jenkins00,lehner03}, but it is generally
mild compared to the high redshift systems \citep{vladilo03}. 
A harder spectrum (i.e. a UV background from QSOs only) would, however,
produce a more discrepant model, e.g., $N($\sii$)$ and $N($\feiii$)$ would
be more difficult to match simultaneously. 

This model implies that the singly-ionized species (\cii, \siii, \mgii, \feii, \Niii, and \sii, 
although the latter may be more subject to ionization correction as illustrated in Fig.~\ref{cloudy-lvc})  
principally trace the neutral gas. The ionization fraction (\hii/H) is 
about 12--20\% (not taken into account the high ions, see below). 
The ``average'' linear size, $L\equiv N({\rm H})/n_{\rm H}$, of the layer of gas in this model is 
$\sim$2--6 kpc, which seems acceptable for a protogalaxy at $z\sim 2.4$
(note that for absorbers with multiple components, $L$ given by Cloudy 
should be treated as an upper limit).
We noted a close kinematic correspondence between 
the ionized and neutral gas is consistent with the picture of an external 
UV source that ionizes the external layer of the neutral gas. 

The Cloudy photoionization simulation, however, fails to produce enough 
\siiv, \civ, \nv\ and \ovi\ (by orders of magnitude for \civ, \nv\ and \ovi\, see Table~\ref{t4a}), 
which implies that these ions must be produced
by other means, such as collisional ionization processes. Such a conclusion was
also drawn by \citet{fox07} in another DLA. We note if 
$\log U \simeq -2.35$, the modeled and observed column densities of \feiii\ and \siiv\
would be similar. This solution would also contribute to a significant amount of \civ, but
would require Al to be deficient by $-0.4$ dex  compared to Si and Fe. Sulfur would be 
underproduced by about 0.2 dex.
Such a solution would also require a very large path length
($\sim$40 kpc) and would nevertheless fail to produce enough \ovi\ and \nv\ columns. 
Therefore this solution is less compelling than the $\log U \sim -3$ solution. 

Collisional equilibrium (CIE) and non-equilibrium collisional ionization (NECI)  may also be important 
sources of ionization, particularly for producing high ions \citep{sutherland93}.
Recently, \citet{gnat07}  have produced new calculations of  CIE and time-dependent
radiative cooling gas as a function of metallicity using updated atomic data.
In galactic environments, highly ionized species may also be produced via
shocks \citep{dopita96}, or in interfaces between cool 
($T\la 10^4$ K) and hot ($T\ga 10^6$ K) gas, e.g., in conductive heating models 
\citep{boehringer87,borkowski90} or turbulent mixing models
\citep{slavin93}. In view of the calculations of \citet{gnat07}, metallicity 
can be an important ingredient; most of the models above assume 
solar metallicity gas. 

To constrain the collisional ionization model, the high-ion column density ratios 
combined with the line-widths of the high-ion components may a priori help us 
to discern between these various models \citep[e.g.,][]{spitzer96}. Unfortunately, 
the results from the profile fitting are somewhat uncertain for the LVC since the 
absorption is so strong (see \S\ref{sec-fit}). From the \siiv\ and \nv\ profiles, 
there are some suggestions of  a mixture of hot ($T \la 3 \times 10^5$ K in 
the $-11$ components) and cool ($T\la 10^4$ K in the $-21$ \km\ component) gas 
(see Table~\ref{t2}, \S\ref{sec-fit}, and Fig.~\ref{spec1}). As we argue above, 
there is also room for an extra component at $\sim 0$ \km, at least in the \siiv\ and 
\civ\ profiles, and hence the components at $-11$ and $+22$ \km\ could 
be narrower. The $N_a(v)$ profiles of \nv\ and \ovi\ show that most of the absorption 
is at $-20 \la v \la +10$ \km\ and in particular the absorption peaks at 
$-20$ and $-11$ \km (see Fig.~\ref{aodp}). This contrasts from the \siiv\ absorption, 
where the $N_a(v)$ profiles peak at $\sim$$-15$ but also at $+22$ \km.  The $N_a(v)$ profile
of \civ\ appears to be somewhat intermediate between \siiv\ on one hand and \nv\ 
and \ovi\ on the other hand. The  $N_a(v)$ profiles of the high ions 
therefore suggest that at $0 < v \la +40$ \km, the highly ionized (in particular \siiv) gas may
be more subject to photoionization, but at  $-40 \la v \la 0$ \km, collisional ionization becomes 
more important. However, we emphasize that both mechanims likely occur at any $v$. 
For the component at $+16$ \km, assuming that a large fraction 
of \siiv\ is photoionized,\footnote{The velocity centroid of \siiv\ is $+22$ \km, 
slightly different from the other high ions, consistent with the idea  that the origins of these ions 
may be different.} the ratios of \civ/\ovi\ and \nv/\ovi\ (taken into account
the intrinsic deficiency of N) in the models of \citet{gnat07} do not fit the observed 
ratios, but models such as conductive interfaces or shock ionizations can produce the observed
high-ion ratios.  For the 
other components, the high-ion ratios are too poorly determined to draw any firm conclusions
about their origin(s), although we note that the line width of the component at $-22$ \km\ implies a very low 
temperature for \civ\ and \siiv\ (see Table~\ref{t2}) and probably for \nv\ (see Fig.~\ref{spec1}), 
suggesting an additional photoionization source (possibly from stars) or some NECI process 
with efficient cooling. 

According to our profile fitting, there is also a close kinematic relationship
between  the high ions and the low ions and atoms. 
A simple explanation for this may involve interface layers between a hot 
($T>10^6$ K) and a cooler ($T\la 10^4$ K) gas. Such interfaces are believed to be 
common in our much evolved Galaxy. However, the column densities of the high ions (in particular \ovi)
in model interfaces are much smaller than observed in individual interfaces \citep{savage06} 
and modeled \citep{borkowski90,slavin93},
requiring (too) many interfaces (and this discrepancy is probably worse in
the DLA, which has a much lower metallicity than the Galaxy). Other interstellar models of high-ion production 
(e.g., radiatively cooling clouds, bubbles blown by supernovae or stellar winds) predict
large high-ion column densities. However, as discussed by \citet{tripp08}, 
the physics of the interfaces may be oversimplified and not so well understood, especially
in the context of high redshift, low metallicity systems. 

From the above, we conclude that
the kinematic and ionization properties of this DLA are consistent  with 
a highly ionized halo that co-rotates with a neutral proto-galactic disk. 
This picture is quite similar to our own Galaxy, which is surrounded by 
a hot highly ionized and a warm ionized halo, revealed by the same high
and intermediate ions as those at high redshifts \citep[e.g.,][]{zsargo03,savage03,howk03}.
However, since several  components are observed, it is also possible that we are
simply seeing  several sheets of neutral gas separated by weakly and highly ionized gas.
In galaxies, a major source for heating and ionizing the interstellar medium is
massive stars. It is well known that these stars inject large amounts of energy in the interstellar
medium  over large scales via their supernovae and stellar winds. This environment may cool and/or 
produce interfaces between cool ($T\la 10^4$ K) and hot  ($T>10^6$ K) gas.
We have shown that the UV background plays an important role in ionizing and heating the present DLA, 
but it cannot explain by itself the origin of the high ions, especially \ovi, \nv, but also \civ, 
and even part of \siiv. Below we argue and show that there is room 
for stars to be present in this DLA-protogalaxy, hence stellar feedback is a possible
origin for the high-ion absorption. 

\subsection{Nature of the neutral gas and star formation}\label{sec-cool}
Above we show that the UV background plays an important role in ionizing and heating 
the present DLA. There is, however, some leeway for the existence of other 
(internal) sources of ionization and heating; e.g., those might help to better model
the column densities of \feiii\ and \siiv. We further investigate the 
nature of the neutral gas in this DLA and the implication for the star formation
rate (SFR) using the limit on \cii*\ absorption and the 
technique described by  \citet{wolfe03} amd \citet{wolfe04}. In this technique, 
we allow the heating and cooling to be balanced in a two phase medium
(cold and warm neutral medium, hereafter CNM and WNM). The heating 
is mostly due to grain photoelectron emission, assuming a Fe depletion 
$[{\rm Fe/Si}] = -0.2$.
The cooling in the CNM and WNM is dominated by  \cii\ at 158 $\mu$m and Ly$\alpha$ emisions, 
respectively \citep[for more details, see][]{wolfe03}. 
The star formation per unit area can be directly related to the photoelectric 
heating rate. The predicted \cii\ 158 $\mu$m cooling rate is compared with that derived
from the strength of the \cii*\ absorption
\citep[e.g.,][]{lehner04}: 
\begin{equation}\label{ecool1}
l_c = 2.89 \times 10^{-20}\frac{N({\rm C}^{+*})}{N({\rm H}^{0})}\, 
	\,{\rm erg}\, {\rm s}^{-1}\,  ({\rm H\,atom})^{-1}\,.
\end{equation}
With the column densities of \cii*\ and \hi\ presented in Table~\ref{t5}, 
we find that for LVC, $\log l_c < -27.8$ (3$\sigma$). 
The upper limits of Wolfe et al. are 2$\sigma$, i.e. 
for our DLA,  $\log  l_c < -28.0$  (2$\sigma$).\footnote{An integration of 
\cii*\ profile over the $[-20,+20]$ \km\ velocity interval would give a 2.5$\sigma$ detection
($W_\lambda = 5.4 \pm 2.2 $ m\AA, $\log N = 12.45\,^{+0.14}_{-0.20}$), which
would imply $l_c = -27.9 \,^{+0.2}_{-0.3}$, consistent with the 2$\sigma$ limit.
We favor the upper limit because the velocity range is not justified in 
view of the other absorption profiles, but it is possible that the value of $l_c$
is not much smaller than the quoted 3$\sigma$ limit.}  This is the lowest known limit
on the cooling rate so far found in a DLA \cite[see the up-to-date sample of][]{wolfe08}. 
Extremely low \cii\ radiative cooling  rates can be found either in cold neutral 
gas where the SFR is low or in warm neutral
gas \citep{wolfire03,wolfe03,lehner04}. Below we show that this is the only known DLA 
where the neutral gas is unambiguously warm and where star formation may occur.

\begin{figure}
\includegraphics[width = 8.5truecm]{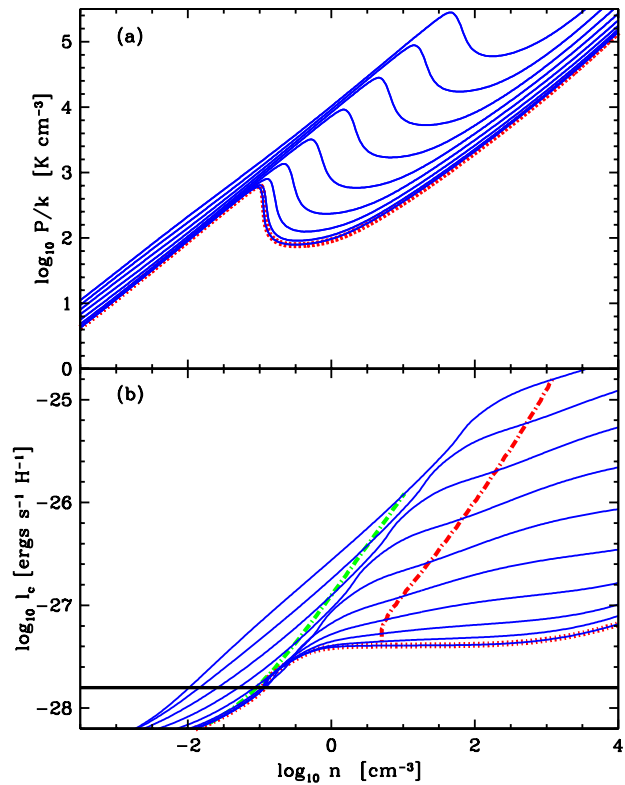}
\caption{The panels show a sequence of equilibrium curves for gas subjected to internal heating rates generated by 
a grid of SFRs per unit area, $\log \Sigma_{\rm SFR}= -4.0$, $-3.5$, $-3.0$, ....0.0 (above the bottom
dotted red curve to top curves) 
in units of M$_\odot$\,yr$^{-1}$\,kpc$^{-2}$.
Extragalactic background radiation is included and the bottom curve shows the result for  background radiation alone.
In panels (a) and (b) we show the pressure and the cooling rate, respectively, against the density of
the gas. In (b) the red dash-dotted curve is the locus of all equilibrium CNM solutions in which 
$P_{\rm eq} = (P_{\rm max}  P_{\rm min})^{0.5}$. The green dash-dotted curve is the same for the WNM solutions.
The solid horizontal line shows the 3$\sigma$ limit derived for the DLA at $z\simeq 2.377$ 
toward J1211+0422.
\label{fig-cool}}
\end{figure}

Fig.~\ref{fig-cool} shows the 
equilibrium solutions for gas subjected to internal and external sources of heating. The internal source
is photoelectric heating by dust grains absorbing starlight produced locally (and an assumed
cosmic-ray contribution), while the external heating is produced
by the extragalactic background from galaxies and QSOs including UV and 
a higher energy component \citep[see][for more details]{wolfe03,wolfe04}. 
We assume no enhancement in $\alpha$-element abundances in estimating the 
dust content \citep[see][]{wolfe04}. 
The red dash-dottedcurve is the locus of all equilibrium CNM solutions
and the green dash-dotted curve is the same for the WNM solutions.
The solid horizontal line shows our 3$\sigma$ limit on $l_c$, which rejects a CNM solution. 
Therefore the neutral gas is warm. The WNM solution also suggests that star formation may 
occur, but only at a rate $\Sigma_{SFR} < 7.1\times 10^{-3}$   M$_\odot$\,yr$^{-1}$\,kpc$^{-2}$. This limit
is about a factor twice larger than the local rate in the Milky Way. 
As we discuss in \S\ref{ssec-metal}, there might be very little dust in this DLA.
We therefore also ran a dust-free model where the only heating is 
by X-ray photoionization and cosmic rays. As in the previous model, their heating  
rates are proportional to the SFRs. The CNM solution is again ruled out and star formation 
is still posible in the WNM solution. A population of DLAs with warm neutral gas and possible 
ongoing star formation exists based on our findings.
Hence many of the sightlines where only  upper limits on $l_c$ are derived may actually probe WNM gas. 

\section{The fully ionized high-velocity gas}\label{sec-ion}
\subsection{Overview}\label{sec-ion-over}
The strong \oi\ $\lambda$1302 transition is completely absent outside 
the velocity range $[-40,+35]$ \km\ (see Fig.~\ref{spec2}). Absorption 
from singly-ionized species with large $\lambda f$ values, \siiii, 
and highly ionized species is seen at higher absolute velocities.
The NHVC, the $+39$ \km\ component,
and the PHVC (see Table~\ref{t2}) have 
$[$O/Si$]\equiv \log N({\rm O^0})/N({\rm Si^+}) - \log ({\rm O/Si})_\odot <-0.8$.
Since Si and O are both $\alpha$-elements (i.e., have similar nucleosynthetic
history), we expect the intrinsic O/Si should be nearly solar. 
The strong underabundance of {\em neutral}\ oxygen compared with
singly ionized silicon implies these clouds are strongly ionized  \citep[see, e.g.,][]{lehner01}.
In all these components, we also find that the column densities of the highly ionized 
species are systematically larger than respective singly-ionized species, 
i.e., $N($\civ$) > N($\cii$)$ and $N($\siiv$) > N($\siii$)$, implying not only hydrogen 
fully ionized in the gas, but also the metals are highly ionized.

While these components could certainly arise in the IGM and be unrelated 
to the DLA at $z \simeq 2.377$, the close velocity proximity to the DLA 
($|v-v_{\rm LVC}| \la 100$ \km) suggests some association, possibly 
via outflows or infalling material on the DLA protogalaxy 
or in form of a highly ionized halo about the protogalaxy that does not 
share the same overall motion. 
Further, ``typical" diffuse IGM conditions would require simultaneously 
$\log n_{\rm H} \la -5$  ($\rho/\bar{\rho} \la 1$) and $[{\rm Z/H}] < -2$ \citep[e.g.,][]{simcoe06}, 
and are unlikely to occur for these absorbers 
since the Hubble 
broadening $b_{\rm H} \approx 220 L({\rm Mpc})$ (where we assume 
a standard cosmology at $z\sim 2.4$) would be larger than the observed line
widths of the individual components.

Following, \citet{fox07}, the escape velocity of a DLA may be estimated if
the virial velocity, $v_{\rm vir}$, of the DLA-protogalaxy is known and assuming 
a spherical geometry: $v_{\rm esc} \approx \sqrt{2} v_{\rm vir}$. 
\citet{haehnelt98} found  $\Delta v_{\rm neut} \approx 0.5$--$0.75 v_{\rm vir} $ \citep[see also,][]{maller01}.
For the DLA presented here $\Delta v_{\rm neut} \simeq 70 $ \km, hence
none of the high-velocity components could escape the gravitational potential 
of the DLA. However, the relation between $v_{\rm vir}$ and $v_{\rm esc}$ 
is not well known and the velocity width and virial velocity, 
while linearly correlated according to cosmological simulations, shows a large scatter around the median value. 
Also, in both \citet{haehnelt98} and \citet{maller01} simulations, the probability distribution of 
$\Delta v_{\rm neut} / v_{\rm vir} $ has a large scatter from $\Delta v_{\rm neut} / v_{\rm vir} \sim 0.2$
to $\sim$1.5, enabling the high-velocity components to be also signatures of feedback-driven flows 
or even an escaping galactic wind from the protogalaxy. It could also be that our line-of-sight passes 
through a highly ionized halo of another galaxy \citep[see, e.g., Fig.~1 in][]{maller03} since
clustering of galaxies near DLAs exists at both low and high redshifts 
\citep[e.g.,][]{chen03,bouche04,cooke06a,cooke06b,ellison07}.
Yet it is unlikely that the fully ionized components probe the core of other galaxies 
(since there is no signature of neutral gas).  

The amount of total hydrogen is unknown in these ionized components,
making it impossible to determine the metallicity. We therefore 
assume that all have a metallicity  similar to the LVC, i.e., we set a priori 
$[{\rm Z/H}] = -1.4$ in our Cloudy simulations. This similarity in the metallicities might arise if  
the high-velocity ionized components have their origins in the
protogalaxy or at least were polluted by the DLA itself or by another protogalaxy
with a similar metallicity.  Metallicity measurements in DLAs at $z \sim 2.4$ show
a large scatter at this redshift with $[{\rm Z/H}]$ ranging from about $-0.5$ dex to $-
2.0$ dex, but with a mean around $-1.4$ dex 
\citep{prochaska03}. Yet if the high-velocity gas was only slightly polluted 
by materials from protogalaxies and is being accreted by the present DLA, a much 
lower metallicity may be expected \citep[e.g., see][]{simcoe04,simcoe06}.

Below we review the properties of each high-velocity components and in particular
explore the possible ionization mechanisms and the connection (if any) between
the various ionized species. Our main aims are to understand (i) if ``simple'' photoionization
and collisional ionization models can reproduce the observed properties (column density
ratios and Doppler broadening) of the ionized gas;
(ii) if \ovi\ and \nv\ {\em necessarily}\ probe {\em hot}\ collisionally ionized gas; (iii) the 
origin(s) of the highly ionized high-velocity components.

\subsection{The negative high-velocity component}\label{sec-NHVC}
For the NHVC, the velocity centroids at $-85$ and $-64$ \km\ are quite similar for
\cii, \alii, \siiv, and \nv\  (see Figs.~\ref{spec1} and \ref{spec2}), 
suggesting some relationship between all these ions. Inspecting the
ratios  of the apparent column densities of various ions with respect to \siiv, 
we find that, in the velocity range 
$[-80,-40]$ \km, there is little or no variation for the low, intermediate,
and high ions  with respect to \siiv, suggesting that the ionization conditions do not change 
much over this velocity range. There is, however, a drop in $N({X^+})/N($\siiv$)$ at 
more negative velocities than $-80$ \km, which also corresponds to an increase of
$N($\nv$)/N($\siiv$)$ and $N($\civ$)/N($\siiv$)$, implying that 
gas becomes more highly ionized at more negative velocities (which is the same 
pattern seen for the LVC). 

In the components at $-100$ and $-85$ \km, the $b$-values  of \siiv\ 
and \civ\ imply a temperature for the gas bearing these ions
of a few times $10^4$ K. In the $-64$ \km\ component, 
higher temperatures may occur, with $T < 1.1\times 10^5$ K. 
At $T<7\times 10^4$ K, collisional ionization models produce 
a fraction of \cii\ larger than the fraction of \civ\  \citep[see][]{gnat07}, 
inconsistent with our observations. Thus
photoionization could play a role for the origin of \civ\
and \siiv\ in components at $-100$ and $-85$ \km. 
Below we therefore model the ionization of the NHVC by producing first a Cloudy 
photoionization simulation that matches the integrated columns of the NHVC, which we 
feel are a more reliable indicator of the average properties of this gas 
than the individual component columns (see \S\ref{sec-fit}).

\begin{table}
\begin{center}
\begin{minipage}{5 truecm}
\caption{Total column densities for the NHVC \label{t5}}
\begin{tabular}{lccc}
\hline
Species   & $\log N$ \\
\hline
\oi\ $\lambda$1302			& $<12.94$   	     \\
\cii\ $\lambda$$\lambda$1036,1334	& $13.69 \pm 0.05$	    	     \\
\civ\ $\lambda$$\lambda$1548,1550	& $>14.66 $   	     \\
\nii\ $\lambda$1083			& $<12.77$   	     \\
\nv\ $\lambda$$\lambda$1238,1242	& $13.27 \pm 0.13$    	     \\
\ovi\ $\lambda$1031			& $>14.80 $	     \\
\alii\ $\lambda$1670			& $11.95 \pm 0.05	$  \\
\aliii\ $\lambda$$\lambda$1854,1862 	& $12.05 \pm 0.08$	\\
\siii\ $\lambda$$\lambda$1260,1304,1524	& $12.67 \pm 0.08$	     \\
\siiii\ $\lambda$1206			& $>13.12$	     \\
\siiv\ $\lambda$1393,1402		& $13.70 \pm 0.02$	     \\
\feiii\ $\lambda$1122			& $< 13.20 $	    \\
\hline
\end{tabular}
Note: Upper limits are 3$\sigma$. 
\end{minipage}
\end{center}
\end{table}

\begin{figure*}
\includegraphics[width = 16truecm]{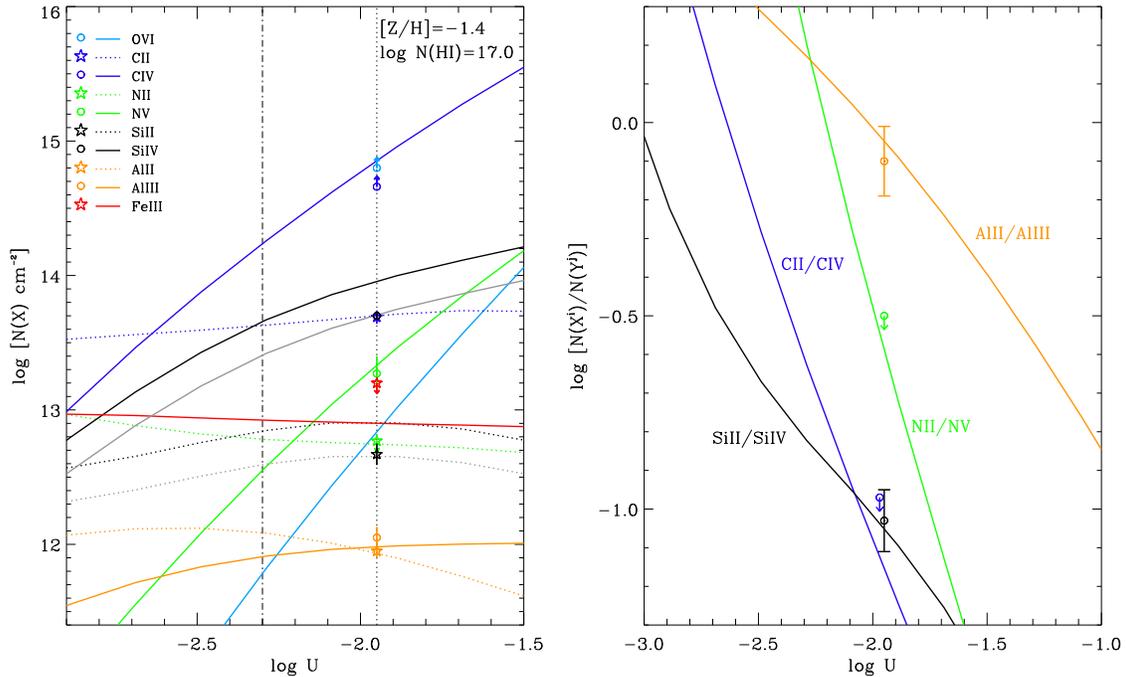}
\caption{{\em Left:} Predicted column densities  for the Cloudy photoionization model of
the NHVC assuming a Haardt-Madau (QSOs+galaxies) spectrum.
Relative solar abundances are assumed.  The various lines show the models for each 
atom or ion (the gray solid and dotted lines for \siiv\ and \siii\ are the Cloudy 
solution $-0.25$ dex, see text for more details), while the symbols are the measurements. 
Symbols with downward arrows are 3$\sigma$ upper limits
and with upward arrows are lower limits. The vertical dotted line corresponds to the 
solution found from the right diagram. The vertical dot-dashed line correspond to 
another possible solution (see text for more details). 
{\em Right:} Ratio of column densities for the observed and models versus the hydrogen 
density. Data are plotted at $\log U = -1.95$ where within about 1$\sigma$ the 
observed ratios of \siii/\siiv\ and \alii/\aliii\ and the limits on \nii/\nv\ and \cii/\civ\ 
agree with the model.  
\label{cloudy-NHVC}}
\end{figure*}

Fig.~\ref{cloudy-NHVC} shows the result of our Cloudy simulation using 
the Haardt \& Madau UV background (see \S\ref{ssec-ion}). In the left panel we 
show the predicted and observed column densities against the density. The metallicity
is a priori assumed to be similar to the LVC.  $N($\hi$)$ is  
chosen so that the model predicts enough column densities for the weakly ionized species. 
It is, however, quite possible, that other (NECI) 
mechanisms may produce both  low and high ionization species (see below).  
We do not show $N($\oi$)$ and $N($\siiii$)$ in Fig.~\ref{cloudy-NHVC} because the estimated
limits are not constraining. 

In the panel on the right-hand side of Fig.~\ref{cloudy-NHVC}, 
we show the column density ratios for multiple 
ionization stages for C, N, Si, and Al against the density. In 
this representation, nucleosynthetic effects do not play any role since
the same elements in different ionization stages are considered. 
We do make the assumption, however, that ions in different ionization
stages are all produced by the same ionization mechanism. 
With those assumptions, the best solution is found at 
$\log U = -1.95$ where the observed ratios of 
\siii/\siiv\ and \alii/\aliii\ and the limits on \nii/\nv\ and \cii/\civ\ 
agree with the model within about 1$\sigma$. 

The vertical dotted line  on the left panel of Fig.~\ref{cloudy-NHVC} shows this solution 
in the column density versus density plot. \civ, \nv, \nii, \alii, and \aliii\ are consistent 
with such a solution (since N may be deficient by $-0.9$ dex, see
\S\ref{sec-prop}, this would require another mechanism to produce \nv, see below), 
but both \siii\ and \siiv\ are over-produced by 0.25 dex. 
The gray dotted and solid lines show the column densities of 
\siii\ and \siiv, respectively, minus 0.25 dex, which fit the data quite well. 
Therefore, if $[$Si/C$] = -0.25$ and $[$Al/C$] = 0$, this photoionization 
model provides a good description for ions with ionization potential less
than about 50 eV. However, this relative abundance pattern is quite
peculiar: neither depletion into dust nor standard nucleosynthesis yields can provide a simple 
explanation for such a relative abundance pattern. If the former plays
a role, Al should also be deficient, while solar or supersolar 
$[$Si/C$]$ are usually observed in low metallicity protogalaxies. 

The assumption that all the ionization stages are produced by 
the same ionizing spectrum could be wrong. For example, if $\log U = -2.3$
(shown in Fig.~\ref{cloudy-NHVC} by the dot-dashed line), 
there is a good   agreement between the model and observation for 
\siiv, \aliii, and \cii\ within about 1$\sigma$. 
We note that this solution matches the \siii/\siiv\ ratio in the velocity
range $[-100,-40]$ \km\ (i.e. the most negative velocity absorption is 
ignored where the gas becomes more highly ionized, see above). 
\civ\ would be under produced, but  \civ\ can be also produced
via collisional ionization (see below). However, with this latter solution,
\alii\ and to lesser extent \siii\ would be overproduced and the \alii/\aliii\ 
ratio does not fit the observations. The ionization potentials 
for \cii, \siii, and \aliii\ are in the range 11.3--24.4 eV, 8.2--16.4 eV, and 5.9--18.8 eV, 
respectively: the discrepancy between the model at $\log U = -2.3$ and 
observations increases as the minimum ionization potentials become smaller. 
It is therefore possible that the model with the current UV background produces too many 
photons at low energies. We note that the typical size for $\log U = -2.3$ 
would be $L \sim 12$ kpc, more reasonable than at  $\log U = -1.95$ where $L \sim 90$ kpc.

It is possible that both the shape and intensity of the UV background 
vary in the Universe. In addition, the escape of ionizing flux from protogalaxies
is uncertain and could change both the shape and intensity of the ionizing 
radiation. The impact of an ``internal" ionizing source could be similar. 
 If the DLA is close to an AGN or a QSO, its ionizing radiation
would have an important impact on the shape and intensity of the UV background. 
We therefore first investigated if a softer ionizing source
would produce a better match to the observations, at least for the low 
ionized species. We considered ionization by a late O-type 
spectrum ($T_{\rm eff} = 35000$ K) with an atmospheric metallicity $[$Z/H$]=-1.4$ 
and an atmosphere from the TLUSTY models \citep{hubeny95}. We find that a model 
with $\log N($\hi$) =17$ fits fairly well the observed column densities of
the singly ionized species \cii, \siii, \alii, and \nii\ (assuming
that N is deficient by at least $-0.7$ dex). 
On the other hand, column densities of the higher ionized gas (including \aliii)
cannot be reproduced by this model, falling short by several orders of magnitude. 
While such a model cannot be sufficient on its own, it is possible that the true ionizing 
spectrum is a combination of soft and hard UV spectra. We therefore considered 
a Haardt \& Madau UV background from QSOs only, i.e. a UV background harder than 
considered previously. A solution with $\log U = -1.35$ ($\log n_{\rm H} = -3.5$) 
can match the column densities of \siiv\ and \nv\ (if nitrogen is deficient by $-1.7$ dex)
and the lower limits on \civ\ and \ovi, while producing negligible \siii\ and \alii. 
But it fails to produce enough \aliii. 
The QSOs only background also produces a solution at $\log U = -2.05$ ($\log n_{\rm H} = -2.8$)
that fits the \siii, \siiv, and \civ\ column densities. This model 
requires the relative abundance of Al to be deficient 
by $-0.6$ dex to match \aliii\ and, and in any case does not produce
enough \civ, \nv, and \ovi. 

We have therefore not found a satisfactory Cloudy photoionization model with a single
source of ionization unless the NHVC has
peculiar abundances or the UV background model 
that includes galaxies and QSOs overestimates the contribution from low energy photons. 
The latter model and explanation ($\log U \simeq -2.3$) appear plausible because
it better fits the \siii/\siiv\ ratio over the velocities  $-90 \la v \la -40$ \km. 
We note that if $[{\rm Z/H}]$ increases in this model,\footnote{To 
first order, reducing/increasing the metallicity 
increases/decreases the hydrogen column density, but the effect is not solely
linear and the shape of the ionization fractions as a function 
of temperature or density can change with $[{\rm Z/H}]$. We also note that
if we adopted $\log N($\hi$) =17 \pm 0.5$ dex (with matching metallicities), 
$\log U$ would only change by about $\pm 0.2$ dex.} the discrepancy between
model and observations for the singly-ionized species  decreases, but  modeled \siii\ and \alii\ column densities 
are  still not in agreement with the observations even at $[{\rm Z/H}] = -0.5$, which is the current 
maximum metallicity observed in a DLA at $z\sim 2.4$. A hard spectrum would seem unlikely to produce the 
observed amount of \nv\ and \ovi\ (and some \civ) because the Hubble flow would
broaden the absorption profiles too much. Therefore, other ionization mechanisms, 
including collisional ionization, may play a role in this gas. 

The observed high-ion ratios,  broadenings of the high ions, and 
$N($\civ$)\gg N($\cii$)$ and $N($\siiv$)\gg N($\siii$)$ are 
incompatible with CIE or NECI models. Other models involving shocks or interfaces 
also require temperatures for the highly ionized gas of a few times $10^5$ K. 
Hence a combination of simple photoionization and collisional ionization models 
cannot reproduce all the observed properties, implying that the assumptions in these models 
are unlikely to hold for the NHVC, especially for the high ions. Therefore physical conditions 
are likely to be more complicated than these simplified scenarios (see \S\ref{sec-comp}).

\subsection{The component at $+39$ \km}\label{sec-cloud39}
\begin{figure}
\includegraphics[width = 8truecm]{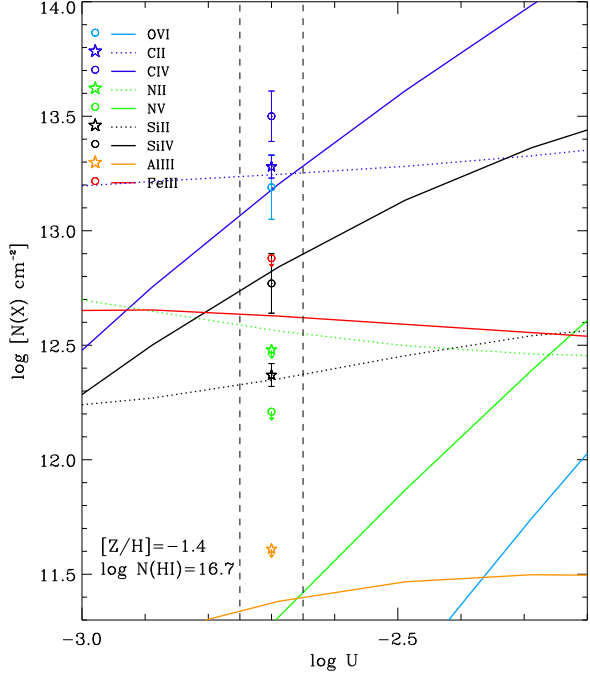}
\caption{Predicted column densities for the Cloudy photoionization model of
$+39$ \km\ component  assuming a Haardt-Madau (galaxies+QSOs) spectrum. 
Relative solar abundances are assumed. 
The various lines show the models for each atoms or ions, while the symbols
are the measurements. Symbols with downward arrows are 3$\sigma$ upper limits,  
upward arrows are lower limits. An adequate solution is found for  $\log U = -2.7$, which 
implies $\log n_{\rm H} = -2.0$, $T = 1.8 \times 10^4$ K,  and $\log N({\rm H}) = \log N({\rm 
H}^+) = 19.4$; i.e. this cloud is fully ionized. The implied path length is $L = 0.8$ kpc.
The most discrepant values to this model are the high ions
(\ovi, \nv, \civ), which likely probe other ionization mechanism(s) 
(see \S\ref{sec-cloud39} for more details). 
\label{cloudy-c39}}
\end{figure}

As for the NHVC, the kinematics of the high, intermediate,  
low ions in the $+39$ \km\ component appear connected (see, e.g., \cii, \siii, \siiv\ in Fig.~\ref{spec2}).
We modeled this component with Cloudy adopting the same assumptions as described in the
previous section. In Table~\ref{t6}, we summarize the column densities for the various
species. The low ions (\siii, \cii) and high ions (\siiv\ and about 80\% of $N($\civ$)$)  
can originate in photoionized gas when $\log n_{\rm H} = -2.0$, 
which implies $\log U = -2.7$, $T = 1.8 \times 10^4$ K,  
and $\log N({\rm H}) = \log N({\rm H}^+) = 19.4$ (see Fig.~\ref{cloudy-c39}). 
The linear size of the absorbing material is about 800 pc. Lower metallicity 
cannot be ruled out based on the broadening of component. 

As for the NHVC, the column densities of \ovi\ and \nv\ (and part of \civ) 
cannot be reproduced by ionization by the UV background from galaxies and QSOs. 
Assuming that the photoionization model is correct, the amount of \civ\ 
not produced by this model would be $\log N($\civ$) \sim 13.2$. With 
this estimate, the ratios of \civ/\ovi\ and \nv/\ovi\ would be compatible
with a CIE model only if N is deficient by at least $-1.2$ dex. However,
this model requires $ T \sim 2 \times 10^5$ K, inconsistent with the $b$-value of \civ\ that 
implies $T < 7 \times 10^4$ K. For \ovi, the (uncertain) 
$b$-value does not suggest temperatures higher than $10^5$ K. It is not clear if other models 
(e.g., conductive interfaces) in low temperature conditions  could reproduce the observed ratios. 
As for the NHVC, we discuss the properties of this component in the context
of feedback models in \S\ref{sec-comp}.

\begin{table}
\begin{center}
\begin{minipage}{5 truecm}
\caption{Column densities of the $+39$ \km\ component \label{t6}}
\begin{tabular}{lccc}
\hline
Species   & $\log N$ \\
\hline
\oi\ $\lambda$1302			& $<12.71$   	     \\
\ovi\ $\lambda$$\lambda$1031,1037	& $13.19 \pm 0.14$	     \\
\cii\ $\lambda$$\lambda$1334		& $13.28 \pm 0.05$	    	     \\
\civ\ $\lambda$$\lambda$1548,1550	& $13.50 \pm 0.11$   	     \\
\nii\ $\lambda$1083			& $<12.48$   	     \\
\nv\ $\lambda$$\lambda$1238		& $<12.21$    	     \\
\aliii\ $\lambda$$\lambda$1854,1862 	& $<11.61$	 \\
\siii\ $\lambda$1193			& $12.37 \pm 0.05$	     \\
\siiv\ $\lambda$1393,1402		& $12.77 \pm 0.13$	     \\
\feiii\ $\lambda$1122			& $<12.88 $  	    \\
\hline
\end{tabular}
Note: Upper limits are 3$\sigma$. 
\end{minipage}
\end{center}
\end{table}

\subsection{The positive high-velocity component}\label{sec-hvc}
With only the detection of \civ, and possibly \ovi\ and \nv, the PHVC ($[+60,+140]$ \km) 
appears even more highly ionized than the gas probed by the NHVC, although the 
column densities of \civ\ are quite similar in both components and the column density of \ovi\ 
appears smaller in the PHVC compared to the NHVC. 
In Table~\ref{t7}, we summarize the measurements for the high ions and the
limits for other strong transitions.  Because $N($\ovi$)$ is uncertain, we assume the 
$N($\civ$)/N($\ovi$)$ ratio is constant over the profile and use $N($\civ) to estimate 
$N($\ovi$)$. This gives the value reported in Table~\ref{t7}. However, for the ionic
ratio used below, we will also take into account the direct measurement $\log N($\ovi$) \le 14.18$
reported in Table~\ref{t1}
(where the lower-equal sign emphasizes that the \ovi\ $\lambda$1037 could be tainted). 

\begin{table}
\begin{center}
\begin{minipage}{5 truecm}
\caption{Total column densities for the PHVC \label{t7}}
\begin{tabular}{lccc}
\hline
Species   & $\log N$ \\
\hline
\cii\ $\lambda$1334			& $<12.63$	    	     \\
\civ\ $\lambda$$\lambda$1548,1550	& $13.68 \pm 0.01$   	     \\
\nii\ $\lambda$1083			& $<12.63$   	     \\
\nv\ $\lambda$1238			& $\le 12.80:$    	     \\
\ovi\ $\lambda$1037			& $\le 13.83:$$^a$	     \\
\siiii\ $\lambda$1206			& $<11.59$	     \\
\siiv\ $\lambda$1393			& $<11.85$	     \\
\hline
\end{tabular}
Note: Upper limits are 3$\sigma$. Because \nv\ is partially 
contaminated, its measurement remains uncertain.  $a$: We report the \ovi\
column density derived from  $N($\ovi$)/N($\civ$) \simeq 1.4$  in the velocity
range $[85,110]$ \km\ (see Figure~\ref{aodhvc}) and  $N($\civ). The ``$\le$'' 
sign emphasizes that the line could be partially contaminated by an unrelated absorber. 
\end{minipage}
\end{center}
\end{table}

The upper limit on the temperature, $T < 1.2 \times 10^5$ K (derived from the 
profile fitting to the \civ\ doublet, see Table~\ref{t2}), could suggest 
that collisional ionization may play a role if thermal motions dominate
the broadening and if the \ovi\ absorption 
is largely contaminated. However,  
at such relatively low temperatures, one would expect a priori 
detection of other ions, in particular \siiv. Models involving 
interfaces (see \S\ref{ssec-ion}) are ruled out because 
even strong transitions of low and intermediate ionized species 
(e.g., \siiii\ $\lambda$1206, \cii\ $\lambda$1334) are not 
detected. The absence of these very strong
transitions is rarely observed in highly ionized gas.  
The high-ion ratios that can help us to discern between ionization processes 
are unfortunately quite uncertain: For \ovi, we have to make 
the dubious assumption that the absorption is not contaminated by some metal line
absorption. The \nv\ absorption is partially blended 
at 1238 \AA\ and completely contaminated at 1242 \AA; therefore only an upper limit can be derived. 
Furthermore, nitrogen is likely to be deficient with respect to the other elements. With these assumptions
and caveats, we find that $N($\civ$)/N($\siiv$) > 68$, $N($\civ$)/N($\ovi$) \ge 0.3$--0.7
(the first value is derived using the total \ovi\ column density 
from the profile fitting -- see Table~\ref{t1}, the second value 
is derived using the AOD column density ratio in the velocity
range $[+85,+110]$ \km), and $N($\civ$)/N($\nv$)\ga 7.7$
(but note that N is likely deficient relative to C). These high-ion ratios with the additional 
constraints on the temperature and the absence of detection 
of low ions are not consistent with any models involving collisional 
ionization.

\begin{figure}
\includegraphics[width = 8truecm]{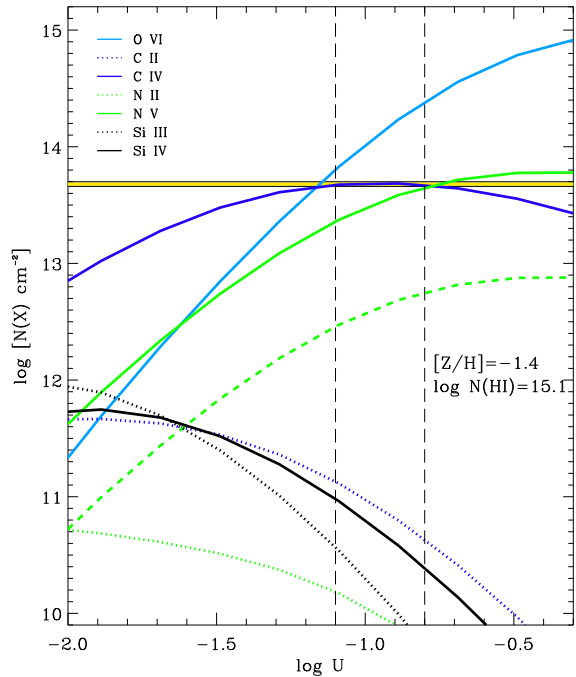}
\caption{Predicted column densities for the Cloudy photoionization model of
the PHVC  assuming a Haardt-Madau (galaxies+QSOs) spectrum. 
Relative solar abundances are assumed. The various lines show the models for 
each atoms or ions. The horizontal yellow line shows the range of observed 
\civ\ column density at 2$\sigma$. Other estimates are reported in Table~\ref{t7}. 
The dashed line represents the solution for \nv\ with $[{\rm N/\alpha}] = -0.9$. 
(see \S\ref{sec-hvc} for more details). 
\label{cloudy-hvc}}
\end{figure}

We have therefore  produced a Cloudy photoionization
simulation with the same assumptions used for the other components, 
in particular that the gas has a metallicity [Z/H$]=-1.4$,
similar to the LVC, and the UV background is from galaxies
and QSOs. In Fig.~\ref{cloudy-hvc}, we show the Cloudy model
that matches the \civ\ column density for  $-1.1\la \log U \la -0.8$. 
Within this range, this model predicts an \ovi\
column density consistent with the range of estimated values, assuming weak
or no blending. The model
produces too much \nv\ when nitrogen has solar relative abundance
with respect to the other metals. However, 
the model is consistent with $N($\nv$)$  if nitrogen is deficient
by $-0.9$ dex (dashed line in Fig.~\ref{cloudy-hvc}). 
All the other limits reported in Table~\ref{t7} are consistent
with the model. Models with $\log U \ll -1.1 $ and $\log U \gg -0.7 $
are rejected because they would contradict the limits on \siiii\ (and \siiv) 
and \ovi, respectively. The path length for this cloud would be quite large
ranging between 50 to 250 kpc. However, if the UV background is dominated by only the QSOs, 
suitable solutions can also be found with smaller $U$, and hence reducing the path length. 
Therefore, despite the absence of low ions, the PHVC does not appear to probe hot 
gas but could represent an extended cool, low density, highly photoionized layer of gas. 

\subsection{Origin(s) of the highly ionized high-velocity gas}\label{sec-comp}
So far we have discussed the highly ionized high-velocity gas in the 
context of simple ionization models. We have shown that photoionization by the UV background
is a possible source for producing the low and intermediate ions, and \siiv\ and part
of \civ. But except for the PHVC, \ovi\ and \nv\ cannot be produced 
by photoionization by the UV background. Photoionization from 
point sources (e.g., from extremely massive stars) is also unlikely as it is not clear that 
such stars can produce enough  $> 113.9$ eV photons to produce the observed
strong absorption in \ovi. It would also require
the highly ionized gas to be extremely close to the ionizing object, 
which is not consistent with the many components observed at both high positive
and negative velocities (we also note that this scenario would have problem 
with the excellent match in the kinematics between low and high ions since the required
large $U$ would make the absorption of the low ions weak). 
For \ovi\ and \nv\ (and in part \civ), 
CIE and NECI models fail to reproduce the 
observed properties (ionic ratios and broadening) of the high ions. 
There is some evidence that the highly ionized gas can be at low temperatures 
based on the $b$-values obtained from the profile fits. The highly ionized gas 
must therefore be relatively dense ($n\ga 0.01$ cm$^{-3}$), or have a relatively
high metallicity, or have a combination of both 
in order to be able to cool to a few times $10^4$ K. Indeed, 
the instantaneous radiative cooling time can be written as \citep[e.g.,][]{gnat07}
 \begin{eqnarray*}
t_{\rm cool}  & = & 219 \left(\frac{T}{10^6\, K}\right) \left(\frac{n}{10^{-3}\, {\rm cm}^{\-3}}\right)^{-1} \\
 	& & \times \left(\frac{\Lambda(T,Z)}{1.3\times 10^{-22}\, {\rm erg\,cm^3\,s^{-1}}}\right)^{-1} \,  {\rm Myr},
 \end{eqnarray*}
where $\Lambda(T,Z)$ is the cooling function, which is dependent on the temperature and the metallicity. 
This expression is valid if the gas is isochoric  
\citep[if instead it is isobaric, 219 should be replaced by 365, see][]{gnat07}. 
For subsolar metallicity ($Z = 0.01 Z_\odot$), and $n = 10^{-3}$ cm$^{-3}$, it would take more than 
the age of the Universe at the absorption epoch ($t_{\rm H}$) for a $10^6$ K gas to cool (if the gas had initially $10^5$ K, 
it would take less than 100 Myr to cool). Therefore, a scenario involving accretion of primordial and low density
gas is not viable if the initial temperature of the hot gas is at least
$10^6$ K since $t_{\rm H} < t_{\rm cool}$. Gas at such high temperatures may exist 
if it was heated, e.g., by supernovae feedback. As the density and the metallicity increase, the
gas can cool more rapidly. 
It is more likely that the highly ionized high-velocity gas is 
a signature of either protogalactic winds/outflows or intra-galaxy, enriched material 
falling onto the protogalaxy. 

At $z>1$, feedback processes such as galactic winds/outflows
or shock heating by supernova remnants are expected to be more common 
because star formation is more active, and these are in fact essential 
for galaxy formation and evolution \citep[e.g.,][]{croton06,mashchenko08}. 
On the other hand galaxy mergers and clustering of galaxies are 
also important at high redshift, favoring the enrichment of gas near 
protogalaxies that may fall onto them or the possibility that a sightline intersects unrelated absorbers 
to the DLA.  \citet{fox07a} favor a scenario 
involving galactic winds because they find high-ion absorption at velocities
larger than the escape velocities of the protogalaxies. However, the knowledge
of $v_{\rm esc}$ in DLAs is not well known (see \S\ref{sec-ion-over}), and we believe there is not
yet enough evidence to favor one scenario over the other one. In principle,
one could use cosmological simulations of protogalaxies that incorporate feedback mechanisms
to try to disentangle these scenarios. 
However,  many aspects of feedback are too complex to reliably model (e.g., metal cooling, 
non-equilibrium ionization effects) and current simulations often lack the spatial and mass resolutions 
to resolve the supernova environment \citep[e.g.,][]{dalla08}. Further
\citet{fangano07} argue it is even difficult to differentiate between outflowing 
and infalling materials in simulations. Nevertheless, we find interesting 
that these recent models can produce low temperatures (a few times 
$10^4$ K) for the highly ionized gas in the galactic outflows, which is so 
difficult to produce in the simplified ionization models used above \citep{oppenheimer06,kawata07,fangano07}.
These models also predict a similarity in the kinematics of 
the low, intermediate, and high ions as observed here.

Local analogs of highly ionized HVCs may also
help to better understand the origin(s) of the highly ionized HVCs at high $z$. 
With the {\em Far Ultraviolet Spectroscopic Explorer (FUSE)}\ 
and the Space Telescope Imaging Spectrograph (STIS) on board of the {\em Hubble Space Telescope},
many highly ionized HVCs (i.e. gas that moves at much different 
velocities than gas associated with the Galactic disk)  were discovered in the halo of the Milky
Way toward extragalactic sightlines \citep{sembach99,lehner02,sembach03,collins04,collins05,fox06}.
At least 60--70\% of the high-latitude sky is in fact covered by \ovi\ HVCs \citep{sembach03}. 
It is  interesting to note that both high and low redshift highly ionized 
HVCs share some common properties: By definition, they are highly ionized, 
but they also often show very complex kinematic and ionization structures, they are often
multiphase and (nearly) fully ionized, and there seems to be some kinematic connection 
between the low and high ions.

As for their high redshift counterparts, the origin of low redshift HVCs remains 
largely enigmatic, and both a ``Galactic" origin  (where the HVCs are the result of feedback processes 
and trace the disk-halo mass exchange, perhaps including the accretion of matter condensing 
from an extended corona) and a ``Local Group" origin (where they are part of the local warm-hot 
intergalactic medium) have been proposed \citep[e.g.,][]{sembach99,nicastro03,gnat04,fox06}. 
Evidence for and against any of these theories are mostly circumstantial, 
although recently \citet{zech07} have discovered two highly ionized HVCs toward 
the star ZNG\,1 in the globular cluster M\,5, firmly demonstrating that some of 
these highly ionized HVCs  originate near the Galactic disk and are signatures 
of Galactic-feedback (the metallicity being supersolar for these HVCs). 
Other highly ionized HVCs associated to galactic 
feedback mechanisms are also found near the LMC \citep{lehner07}. 
On the other hand, some HVCs are also clearly related to an accretion phenomenon or tidal products  
some HVCs are also related to an accretion phenomenon or tidal products 
\citep[e.g., Complex C and Magellanic Stream,][]{tripp03,sembach03}.
Therefore, while an IGM 
origin cannot be ruled out for some HVCs, several highly ionized HVCs in the 
low redshift Universe have been associated with galactic feedback and accretion processes. 
At high redshift we noted that the column densities of the high ions are far too 
large for these ions to arise in the IGM. 
At both high and low redshifts, highly ionized HVCs may therefore be connected to 
gas outflowing from  the host protogalaxy or galaxy. For example, \citet{simcoe06}  presented 
evidence of chemically enriched high-velocity \ovi\ and \civ\ absorbers 
near galaxies at $z\sim 2.3$ with possible high [Si/C] suggestive of 
supernovae type II enrichment, and therefore favoring outflowing matter. 
For the present DLA, there is not enough evidence to favor feedback over
accretion.

While we listed similar properties for the highly ionized HVCs at low and high $z$, they
also have differences. At high $z$, \civ\ (IP: 47.9--64.5 eV) appears to trace
both photoionized and collisionally ionized gas, much like
\siiv\ (IP: 33.5--45.1 eV) at low redshift. Narrow components in \nv\ and \ovi\ absorption 
(and \civ\ unexplained by photoionization by the UV background)  
are observed at high $z$, but are relatively rare or absent at low redshift 
(although part of this could be an instrumental bias since the spectra of the \ovi\ HVCs at $z\sim 0$ 
have $R \simeq 20,000$ compared to 45,000 for the present observations -- e.g., Savage, Sembach, \& Howk
2001 reported a narrow \nv\ component with $b = 7 \pm 2$  \km\ in the Milky Way gas). 
The high-velocity absorption near DLAs is often much stronger than near
our Galaxy and the high ion ratios appear to vary much more in DLAs (at least in the present DLA) 
than in or near the Galaxy. The change in the ionizing background spectral energy distribution 
may explain some of these differences for \siiv\ and part of \civ. 
However, as we have demonstrated here, many of the highly ionized high-velocity 
components  (\ovi, \nv, and some \civ) in the DLA cannot be explained by 
photoionization alone. Therefore some of the listed differences above between 
the highly ionized high-velocity gas at low and high $z$ also suggest an evolution in 
the properties of feedback processes or the environment near the (proto)galaxies.

\section{Discussion}\label{sec-disc}
Before summarizing our main findings, we argue that this DLA is typical in many regards. 
From a standpoint of the metallicity ($[{\rm Z/H}] = -1.41 \pm 0.08$), the main 
component of the DLA at $z=2.377$ toward J1211+0422 is actually quite ordinary  
\citep[see, e.g.,][for a summary of the metallicity evolution of DLAs]{prochaska03}. 
The full velocity width of the neutral absorption of about 70 \km\ is also not 
far from the median value of 90 \km\ derived in a large sample of DLAs \citep{prochaska97}. 
The $N($\hi$)$ value lies near the median of the DLA \hi\ column density distribution.
The $\alpha$-element enhancement relative to Fe-peak elements and N is also not 
out of the ordinary \citep{lu96,prochaska02,petitjean08}. But what may 
really distinguish this DLA from others is the absence of cold neutral gas signature. However,
$32/76 = 42$\% of the DLAs studied by \citet{wolfe08} have no detectable \cii*\ absorption, 
possibly indicating the neutral gas is warm. None of the upper limits derived on $l_c$ are as stringent
as in the present DLA and improving the signal in the spectra of these DLAs would help to understand
if the population of DLAs with WNM is important.

We also saw that the high, intermediate, and low ions in 
this DLA seem to follow each other quite well kinematically. 
\citet{wolfe00a} found 
that \civ\ and low ions rarely follow the same kinematic distribution,
\siiv\ and \civ\ appear strongly correlated, and in many cases the 
\aliii\ profiles are better correlated with those of the low ions than with those 
of \civ\ and \siiv\ \citep[see also][]{lu96}. However, even though
the components of the high-ion profiles are generally disjoint from the 
low-ion profiles, \citet{wolfe00a} noted that the high-ion
and low-ion profiles are roughly centered on similar velocities and 
that the high-ion profiles have a full-width systematically similar
to or larger than that low-ion profiles, properties similar to the present 
DLA \citep[see also][]{fox07a,fox07}.  The ionized  DLA described 
by \citet{prochaska02a}  shows 
some correlation between the velocity centroids of the low, intermediate, and 
high ions. 
The absence of clear kinematic alignment between the high, intermediate, and low-ion profiles
in every DLA may, however, not be entirely surprising. In our own Galaxy, there is a significant
variety in the type of associations between the various phases of the interstellar gas,  
which are generally attributed to sightline projection effects and the physical processes 
that give rise to the high-ions. 

We also found that the amount of ionized gas associated to this DLA is very large. 
To gauge the importance of the highly ionized gas relative to the neutral gas 
we can estimate the amount of hydrogen in the highly ionized phase from 
$N($\hii$) = N($\nv$)/(f_{\rm N\,V} \, ({\rm N/H})_\odot 10^{[{\rm N/H}]})$, where $f_{\rm 
N\,V} = N($\nv$)/N({\rm N}) \simeq 0.2$ 
is the ionization fraction of N at $T \simeq 2 \times 10^5$ K
when using the NECI calculations from \citet{gnat07}. 
At $T$ greater or lower than $2 \times 10^5$ K, $f_{\rm N\,V}< 0.2$.
We assume that the N abundance is the same in the neutral 
and ionized phase,  so that $[{\rm N/H}] = [$\nni/\hi$] \simeq -2.3$. 
For the LVC, $\log N($\hi$) = 20.8$, $\log N($\hii$) \sim 20.5$ (highly ionized gas)
and  $\log N($\hii$) \sim 20.1$ (photoionized gas).
For the NHVC, $\log N($\hii$) \sim 20.5$ (highly ionized gas)
and  $\log N($\hii$) \sim 20.2$ (photoionized gas). Using the lower limits
on $N($\ovi$)$ (Tables~\ref{t4a} and \ref{t5}), we find consistent \hii\ amounts
for the highly ionized gas with $\log N($\hii$) > 20.3$
in the LVC and NHVC (in that case  $[{\rm O/H}] =  -1.4$ and $f_{\rm O\,VI}  \la 0.2$).
The $+39$ \km\ component and PHVC are negligible with 
respect to hydrogen amounts in the LVC and NHVC. 
Summing the various components, $N_{\rm tot}($\hii$)/N($\hi$) \ga 1.5$, 
where $N_{\rm tot}($\hii$)$ includes contribution from both the photoionized gas and 
highly ionized gas. As we discuss above there is some uncertainty 
in the processes that give rise to \ovi\ and \nv. Nevertheless the ionization
fraction of these species is unlikely to be much more than 0.2 based 
on CIE/NECI and photoionization models. The metallicity
correction in the high-velocity components is uncertain because no direct metallicity
estimates can be done. If the metallicity is not similar to the DLA
itself, it is more likely to decrease in these components, although
if there is enough pollution from other nearby protogalaxies an increase
may be possible. The limit  $N_{\rm tot}($\hii$)/N($\hi$) \ga 1.5$ is larger than those typically 
derived by \citet{fox07} for DLAs with similar total \hi\ column density, but is comparable to 
ionized-to-neutral ratios derived in DLAs with lower $N($\hi$)$ and sub-DLAs 
\citep{fox07,fox07c}. The  use of \nv\  may have allowed us to simply derive a more stringent lower limit.

Finally, we observe that \citet{fox07} argue that the highly ionized gas
may help to solve (in part) the missing metal ``problem" at high redshift \citep{pettini99} since if the 
\ovi-bearing gas has $T\sim 10^6$ K, then  $f_{\rm O\,VI} \sim 0.003 $ (according to CIE models), which
would imply  that the \ovi\ phase has enough metals to solve the issue. 
In our analysis, several components have $T \la 3\times 10^5$ K, consistent
with $f_{\rm O\,VI} \la 0.2$. However, there is also evidence for \ovi-bearing gas with $T<9\times 10^4$ K, 
which would also imply $f_{\rm O\,VI} \ll 0.2$ if CIE/NECI applies. But CIE/NECI (and photoionization) provides a 
poor description of this gas.
To quantify accurately the importance of the highly ionized phase in DLAs requires first  
a better understanding of the high-ion absorption in a large and (relatively) uncontaminated sample
(with the purpose of obtaining a statistical sample where the distribution of $b$ in the {\em individual}\
components could be studied), the ionization fractions of the high ions (requiring models with more realistic physics), 
and metallicities in the ionized plasma (which is likely to rely on similar assumptions made in
the present work).

\section{Summary}\label{sec-sum}
We have presented Keck HIRES observations of the absorbing material associated with a DLA at 
$z\simeq 2.377$  along the  QSO J1211+0422 ($z_{\rm em} = 2.5412$) sightline with the goal of 
exploring the properties of the highly ionized gas, its connection with the neutral gas, its origin(s), 
and its possible relationship to feedback processes.
The absorbing material at $z\simeq 2.377$ has little contamination, especially 
in key species, such as \siiv, \civ, and \nv, which allows us to undertake detailed
modeling of the high-ion profiles, to assess their relationship with low ions and study
the ionization mechanisms at play. 

The main results of our analysis for the gas that has velocities in the range $[-40,+35]$ 
\km\ interval where the neutral absorbing material is confined are as follows: 

1. Where there is neutral absorbing material, intermediate- and high-ion absorption 
is also observed.  The profiles of the low and high ions have several 
components. The profile fits to the high ions show that there is a close kinematic relationship 
between the high and low ions. 

2. The metallicity of this DLA, $[{\rm Z/H}] = -1.41 \pm 0.08$,  is similar to 
the average metallicity of other DLAs at $z\sim 2.4$. With $[{\rm Si/S}] = +0.02 \pm 0.06$, 
there is little evidence for silicate grains. The low ratio 
of $[{\rm N/Si}] = -0.88 \pm 0.07$ implies that nucleosynthesis plays 
a role in the observed relative abundances. Hence, the enhancement of 
$[{\rm Si/Fe}] = +0.23 \pm 0.05$ could be solely due to nucleosynthesis. However, 
within the uncertainties, a mixture of dust depletion of Fe and Ni ($[{\rm Si/Ni}]\simeq [{\rm Si/Fe}]$) 
and nucleosynthesis is a possibility.

3. A Cloudy simulation, where the source 
of the photoionization is dominated by the Haardt \& Madau galaxies plus QSOs spectrum, 
reproduces well the column densities of the singly- and doubly-ionized species. 
Within this simulation, about 85\% of the gas is  neutral  and the gas is warm ($T\sim 10^4$ K). 
However, this model fails completely to produce large enough high-ion columns (especially for \civ, \nv, and \ovi). 
Possible models for the production of the high ions include conductive interfaces
or shocks. Including the highly ionized gas, the column of ionized gas in this 
velocity range is similar to the amount of the neutral gas, with
$N($\hii$)/N($\hi$) \ga 0.8$. 

4. We derive the lowest upper limit on \cii*\ in a DLA to date, 
which implies a cooling rate $l_c < 10^{-27.8}$ erg\,s$^{-1}$ per H atom at 3$\sigma$. 
Using the models of \citet{wolfe03} and assuming a Fe dust depletion of  $[{\rm Fe/Si}] = -0.2$, 
we show that the neutral gas can only be warm. We also show that the star formation rate is
$< 7.1\times 10^{-3}$   M$_\odot$\,yr$^{-1}$\,kpc$^{-2}$, i.e. we cannot rule out 
that star formation may occur at a rate about twice the local rate of the Milky Way.  
These conclusions hold even if the environment is essentially free of dust. 
DLAs may therefore have warm neutral gas and ongoing star formation at the same time. 
Improved upper limits on $N($\cii*$)/N($\hi$)$ in DLAs may help to discern the importance of this population
of warm neutral absorbers. 

5. This DLA is made up of warm neutral and warm (and possibly hot) ionized gas
that are comoving with similar velocities given the close
kinematic relationship between the low and high ions. 
Despite the similar velocities for all the species, the DLA has multiple
phases. A picture where the neutral gas is surrounded by a highly ionized halo is plausible,
but not unique. Because several  components are observed, it is also possible that we are seeing 
several sheets of neutral gas separated by weakly and highly ionized gas.

For gas outside of the LVC absorption, i.e., at 
$|v| \ga +39$ \km, absorption is only observed in singly-ionized and higher ionized 
species (i.e. there is no signature of neutral gas). The amount of gas
in the highly ionized phase is larger than the amount in the weakly ionized phase. 
Our main results for the high-velocity gas are as follows: 

6. In the velocity range $[+80,+120]$ \km, {\em only}\ \civ\ and  \ovi\
absorption is detected. This gas has one of the highest known \civ/\siiv\ ratios,
with $N($\civ$)/N($\siiv$)>68$. Other ions with very strong transitions, such as \cii\ and
\siiii, are totally absent. Paradoxically, the observed properties ($N, b$) are better
explained with a cool, photoionized model than a model involving collisional ionization
(in or out of equilibrium). 

7. For the components at $+39$ \km\ and in the velocity range $[-120, -40]$ \km,
low and high ions are detected and appear to be kinematically 
related. A photoionization model using the Haardt \& Madau (galaxies+QSOs) spectrum
can reproduce the column densities of the ionized species in the $+39$ \km\ 
component, except those of \ovi\ and \nv. For the gas in the velocity range $[-120, -40]$ \km,
Cloudy results are difficult to reconcile with the observations without invoking
peculiar relative abundances (subsolar $[{\rm C/Si}]$ and solar $[{\rm C/Al}]$) 
or a different spectral energy distribution for the ionizing spectrum. In any case, photoionization 
does not appear to be a major source for \ovi\ and \nv. The broadening of the profiles implies temperatures 
that are too low to produce \nv\ and \ovi\ (and in part \civ) in CIE (e.g., at $-100$ \km, $b($\ovi$) = 9.5 \pm 1.1$ \km, 
implying $T <10^5$ K; in several components of \civ, $b($\civ$) \approx 4$--10 \km, implying $T\la 1$--$7\times 10^4$ K).
This indicates that cooling must be important and efficient, hence requiring the gas
to be relatively dense ($n_{\rm H} > 10^{-3}$ cm$^{-3}$) and/or rich in metals. 

Combining our results with other works \citep[in particular][]{fox07}, 
we  make some more general conclusions concerning the high ions observed in high redshift DLAs: 

8. In the present DLA, \siiv\  probes predominantly photoionized gas, while \civ\ has the most 
ambiguous origin. Photoionization by a hard spectral source (e.g., QSOs) can produce 
significant \civ\ absorption but is generally not sufficient to explain the total observed column. 
In our Galaxy, \siiv\ is the ``ambiguous'' high ion. The difference in the spectral energy distribution
of the ionizing background flux between the high and low $z$ Universe 
may explain this difference for the most part. In view of other \siiv\ and \civ\ studies in DLAs, 
these conclusions are likely to hold for the whole population of DLAs. 

9. Combining our results with those of \citet{fox07}, DLAs can have strong \ovi\ 
absorption with in some cases \nv\ absorption. In our DLA and for another DLA in 
\citet{fox07}, simple photoionization models fail to reproduce enough columns for these ions. 
However, these ions do not necessarily probe hot ($T> 2\times 10^5$ K) gas, i.e. their mere presence
does not necessarily imply the presence of hot gas. Non-thermal, turbulent motions may broaden 
the high-ion profiles, and for the present DLA, there is evidence for narrow ($T <10^5$ K) 
components not associated with photoionized gas. We also find that  
simple CIE and non-equilibrium radiative cooling ionization models  fail for these
ions and part of \civ\  when the line-widths are available
(i.e. the models cannot reproduce simultaneously the high-ion ratios  and  $b$-values). 
Both collisional and photoionization models are likely more complex than those presented here.
However, in view of our results and those of \citet{fox07}, we believe it is safe to conclude that 
\nv\ and \ovi\ (and part of \civ) associated to DLAs must 
exist in a separate phase than the photoionized phase. In that sense,
they appear better tracers of feedback (or accretion) processes in high redshift DLAs than \siiv\ and \civ.

Future detailed analysis of \ovi\ and \nv\ absorption in a relatively uncontaminated and larger sample
than presently published would be extremely useful, in particular to constrain the $b$-value distribution 
of the individual components. \civ\ and \siiv\ should not be dismissed since they
are far less contaminated by the Ly$\alpha$ forest. 
A detailed profile fitting on a large sample of \civ\ and \siiv\ absorbers associated to DLAs
in order to extract the individual $b$-values would be very valuable since \siiv\ mostly 
probes photoionized gas while \civ\ probes both photoionized and collisionally ionized gas.
This work provides a step in that direction and we 
plan to further investigate the nature of the highly ionized gas in DLAs in the future.

\section*{acknowledgements}

The authors wish to recognize and acknowledge the very significant cultural 
role and reverence that the summit of Mauna Kea has always had within the 
indigenous Hawaiian community. We are most fortunate to have the opportunity 
to conduct observations from this mountain. We also acknowledge the Keck support 
staff for their efforts in performing these observations. This research has made use of 
the NASA's Astrophysics Data System Abstract Service, the SIMBAD database,
operated at CDS, Strasbourg, France, the NASA/IPAC Extragalactic Database (NED) 
operated by the Jet Propulsion Laboratory, California Institute of Technology, 
under contract with the NASA, and SDSS data.

\bsp

\label{lastpage}


\begin{thebibliography}{}
\bibitem[\protect\citeauthoryear{Aguirre et al.}{2001}]{aguirre01} 
Aguirre A., Hernquist L., Schaye J., Katz N., Weinberg D.~H., Gardner J., 2001, ApJ, 
561, 521 

\bibitem[\protect\citeauthoryear{Asplund, Grevesse, \& Sauval}{2006}]{asplund06} 
Asplund M., Grevesse N., Sauval A.~J., 2006, CoAst, 147, 76 

\bibitem[\protect\citeauthoryear{Bland-Hawthorn \& Cohen}{2003}]{bland03} 
Bland-Hawthorn J., Cohen M., 2003, ApJ, 582, 246 

\bibitem[\protect\citeauthoryear{Boehringer \& Hartquist}{1987}]{boehringer87} 
Boehringer H., Hartquist T.~W., 1987, MNRAS, 228, 915 

\bibitem[\protect\citeauthoryear{Borkowski, Balbus, \& Fristrom}{1990}]{borkowski90} 
Borkowski K.~J., Balbus S.~A., Fristrom C.~C., 1990, ApJ, 355, 501 

\bibitem[\protect\citeauthoryear{Bouch{\'e} \& Lowenthal}{2004}]{bouche04} 
Bouch{\'e} N., Lowenthal J.~D., 2004, ApJ, 609, 513 

\bibitem[\protect\citeauthoryear{Chen \& Lanzetta}{2003}]{chen03} 
Chen H.-W., Lanzetta K.~M., 2003, ApJ, 597, 706 

\bibitem[\protect\citeauthoryear{Collins, Shull, \& Giroux}{2004}]{collins04} 
Collins J.~A., Shull J.~M., Giroux M.~L., 2004, ApJ, 605, 216 

\bibitem[\protect\citeauthoryear{Collins, Shull, \& Giroux}{2005}]{collins05} 
Collins J.~A., Shull J.~M., Giroux M.~L., 2005, ApJ, 623, 196 

\bibitem[\protect\citeauthoryear{Cooke et al.}{2006a}]{cooke06a} 
Cooke J., Wolfe A.~M., Gawiser E., Prochaska J.~X., 2006a, ApJ, 652, 994 

\bibitem[\protect\citeauthoryear{Cooke et al.}{2006b}]{cooke06b} 
Cooke J., Wolfe A.~M., Gawiser E., Prochaska J.~X., 2006b, ApJ, 636, L9 

\bibitem[\protect\citeauthoryear{Croton et al.}{2006}]{croton06} 
Croton D.~J., et al. 2006, MNRAS, 365, 11

\bibitem[\protect\citeauthoryear{Dalla Vecchia \& Schaye}{2008}]{dalla08} 
Dalla Vecchia C., Schaye J., 2008, MNRAS, submitted, [arXiv:0801.2770]

\bibitem[\protect\citeauthoryear{Dopita \& Sutherland}{1996}]{dopita96} 
Dopita M.~A., Sutherland R.~S., 1996, ApJS, 102, 161 

\bibitem[\protect\citeauthoryear{Ellison et al.}{2007}]{ellison07} 
Ellison S.~L., Hennawi J.~F., Martin C.~L., Sommer-Larsen J., 2007, MNRAS, 378, 801 

\bibitem[\protect\citeauthoryear{Fangano, Ferrara, \& Richter}{2007}]{fangano07} 
Fangano A.~P.~M., Ferrara A., Richter P., 2007, MNRAS, 381, 469 

\bibitem[\protect\citeauthoryear{Ferland et al.}{1998}]{ferland98} 
Ferland, G.~J., Korista, K.~T., Verner, D.~A., Ferguson, J.~W., 
Kingdon, J.~B., \& Verner, E.~M., 1998, PASP, 110, 761 

\bibitem[\protect\citeauthoryear{Fitzpatrick \& Spitzer}{1997}]{fitzpatrick97}
Fitzpatrick, E.L., Spitzer, L. 1997, ApJ, 475, 623

\bibitem[\protect\citeauthoryear{Fox et al.}{2007a}]{fox07a} 
Fox A.~J., Ledoux C., Petitjean P., Srianand R., 2007a, A\&A, 473, 791 

\bibitem[\protect\citeauthoryear{Fox et al.}{2007b}]{fox07} 
Fox A.~J., Petitjean P., Ledoux C., Srianand R., 2007b, A\&A, 465, 171 

\bibitem[\protect\citeauthoryear{Fox et al.}{2007c}]{fox07c} 
Fox A.~J., Petitjean P., Ledoux C., Srianand R., 2007c, ApJ, 668, L15 

\bibitem[\protect\citeauthoryear{Fox, Savage, \& Wakker}{2006}]{fox06} 
Fox A.~J., Savage B.~D., Wakker B.~P., 2006, ApJS, 165, 229 

\bibitem[\protect\citeauthoryear{Gnat \& Sternberg}{2004}]{gnat04} 
Gnat O., Sternberg A., 2004, ApJ, 608, 229 

\bibitem[\protect\citeauthoryear{Gnat \& Sternberg}{2007}]{gnat07} 
Gnat O., Sternberg A., 2007, ApJS, 168, 213 

\bibitem[\protect\citeauthoryear{Haehnelt, Steinmetz, \& Rauch}{1998}]{haehnelt98} 
Haehnelt M.~G., Steinmetz M., Rauch M., 1998, ApJ, 495, 647 

\bibitem[\protect\citeauthoryear{Heckman et al.}{2000}]{heckman00} 
Heckman T.~M., Lehnert M.~D., Strickland D.~K., Armus L., 2000, ApJS, 129, 493 

\bibitem[\protect\citeauthoryear{Henry, Edmunds, K{\"o}ppen}{2000}]{henry00} 
Henry R.~B.~C., Edmunds M.~G., K{\"o}ppen J., 2000, ApJ, 541, 660 

\bibitem[\protect\citeauthoryear{Henry, Nava, \& Prochaska}{2006}]{henry06} 
Henry R.~B.~C., Nava A., Prochaska J.~X., 2006, ApJ, 647, 984 

\bibitem[\protect\citeauthoryear{Henry \& Prochaska}{2007}]{henry07} 
Henry R.~B.~C., Prochaska J.~X., 2007, PASP, 119, 962 

\bibitem[\protect\citeauthoryear{Herbert-Fort et al.}{2006}]{herbert06} 
Herbert-Fort S., Prochaska J.~X., Dessauges-Zavadsky M., Ellison S.~L., 
Howk J.~C., Wolfe A.~M., Prochter G.~E., 2006, PASP, 118, 1077 

\bibitem[\protect\citeauthoryear{Howk, Sembach, \& Savage}{2003}]{howk03} 
Howk J.~C., Sembach K.~R., Savage B.~D., 2003, ApJ, 586, 249 

\bibitem[\protect\citeauthoryear{Howk, Wolfe, \& Prochaska}{2005}]{howk05} 
Howk J.~C., Wolfe A.~M., Prochaska J.~X., 2005, ApJ, 622, L81 

\bibitem[\protect\citeauthoryear{Hubeny \& Lanz}{1995}]{hubeny95} 
Hubeny I., Lanz T., 1995, ApJ, 439, 875 

\bibitem[\protect\citeauthoryear{Jenkins et al.}{2000}]{jenkins00} 
Jenkins E.~B., et al., 2000, ApJ, 538, L81 

\bibitem[\protect\citeauthoryear{Jenkins \& Tripp}{2006}]{jenkins06} 
Jenkins E.~B., Tripp T.~M., 2006, ApJ, 637, 548 

\bibitem[\protect\citeauthoryear{Kawata \& Rauch}{2007}]{kawata07} 
Kawata D., Rauch M., 2007, ApJ, 663, 38 

\bibitem[\protect\citeauthoryear{Kirkman \& Tytler}{1997}]{kirkman97} 
Kirkman D., Tytler D., 1997, ApJ, 484, 672 

\bibitem[\protect\citeauthoryear{Ledoux, Petitjean, \& Srianand}{2003}]{ledoux03} 
Ledoux C., Petitjean P., Srianand R., 2003, MNRAS, 346, 209 

\bibitem[\protect\citeauthoryear{Lehner}{2002}]{lehner02} 
Lehner N., 2002, ApJ, 578, 126 

\bibitem[\protect\citeauthoryear{Lehner \& Howk}{2007}]{lehner07} 
Lehner N., Howk J.~C., 2007, MNRAS, 377, 687 

\bibitem[\protect\citeauthoryear{Lehner et al.}{2003}]{lehner03} 
Lehner N., Jenkins E.~B., Gry C., Moos H.~W., Chayer P., Lacour S., 2003, 
ApJ, 595, 858 

\bibitem[\protect\citeauthoryear{Lehner, Keenan, \& Sembach}{2001}]{lehner01} 
Lehner N., Keenan F.~P., Sembach K.~R., 2001, MNRAS, 323, 904 

\bibitem[\protect\citeauthoryear{Lehner, Wakker, \& Savage}{2004}]{lehner04} 
Lehner N., Wakker B.~P., Savage B.~D., 2004, ApJ, 615, 767 

\bibitem[\protect\citeauthoryear{Lu et al.}{1996}]{lu96} 
Lu L., Sargent W.~L.~W., Barlow T.~A., Churchill C.~W., Vogt S.~S., 1996, 
ApJS, 107, 475 

\bibitem[\protect\citeauthoryear{Lu, Sargent, \& Barlow}{1998}]{lu98} 
Lu L., Sargent W. L. W., Barlow T. A., 1998, AJ, 115, 55

\bibitem[\protect\citeauthoryear{Maller et al.}{2001}]{maller01} 
Maller A.~H., Prochaska J.~X., Somerville R.~S., Primack J.~R., 2001, 
MNRAS, 326, 1475 

\bibitem[\protect\citeauthoryear{Maller et al.}{2003}]{maller03} 
Maller A.~H., Prochaska J.~X., Somerville R.~S., Primack J.~R., 2003, 
MNRAS, 343, 268 

\bibitem[\protect\citeauthoryear{Martin}{2006}]{martin06} 
Martin C.~L., 2006, ApJ, 647, 222 

\bibitem[\protect\citeauthoryear{Mashchenko, Wadsley, \& Couchman}{2008}]{mashchenko08} 
Mashchenko S., Wadsley J., Couchman H.~M.~P., 2008, Sci, 319, 174 

\bibitem[\protect\citeauthoryear{Matteucci}{2001}]{matteucci01} 
Matteucci F., 2001, ASSL, 253

\bibitem[\protect\citeauthoryear{Meiksin}{2008}]{meiksin08} 
Meiksin A.~A., 2008, Reviews of Modern Physics, submitted [arXiv:0711.3358]

\bibitem[\protect\citeauthoryear{Morton}{2003}]{morton03}
Morton, D. C.  2003, ApJS, 149, 205

\bibitem[\protect\citeauthoryear{Nicastro et al.}{2003}]{nicastro03} 
Nicastro F., et al., 2003, Natur, 421, 719 

\bibitem[\protect\citeauthoryear{Nissen et al.}{2007}]{nissen07} 
Nissen P.~E., Akerman C., Asplund M., Fabbian D., Kerber F., Kaufl H.~U., Pettini M., 2007, A\&A, 469, 3

\bibitem[\protect\citeauthoryear{Noterdaeme et al.}{2008}]{noterdaeme08} 
Noterdaeme P., Ledoux C., Petitjean P., Srianand R., 2008, A\&A, 481, 327

\bibitem[\protect\citeauthoryear{Oppenheimer \& Dav{\'e}}{2006}]{oppenheimer06} 
Oppenheimer B.~D., Dav{\'e} R., 2006, MNRAS, 373, 1265 

\bibitem[\protect\citeauthoryear{Petitjean, Ledoux, \& Srianand}{2008}]{petitjean08} 
Petitjean P., Ledoux C., Srianand R., 2008, A\&A, 480, 349 

\bibitem[\protect\citeauthoryear{Pettini}{1999}]{pettini99} 
Pettini M., 1999, cezh.conf, 233 

\bibitem[\protect\citeauthoryear{Pettini et al.}{1999}]{pettini99a} 
Pettini M., Ellison S.~L., Steidel C.~C., Bowen D.~V., 1999, ApJ, 510, 576 

\bibitem[\protect\citeauthoryear{Pettini et al.}{2001}]{pettini01} 
Pettini M., Shapley A.~E., Steidel C.~C., Cuby J.-G., Dickinson M., 
Moorwood A.~F.~M., Adelberger K.~L., Giavalisco M., 2001, ApJ, 554, 981 

\bibitem[\protect\citeauthoryear{Prochaska et al.}{2002a}]{prochaska02a} 
Prochaska J.~X., Henry R.~B.~C., O'Meara J.~M., Tytler D., Wolfe A.~M., Kirkman D., 
Lubin D., Suzuki N., 2002a, PASP, 114, 933 

\bibitem[\protect\citeauthoryear{Prochaska et al.}{2003}]{prochaska03} 
Prochaska J.~X., Gawiser E., Wolfe A.~M., Castro S., Djorkovski S.~G., 2003, ApJ, 595, L9 

\bibitem[\protect\citeauthoryear{Prochaska et al.}{2002b}]{prochaska02} 
Prochaska J.~X., Howk J.~C., O'Meara  J.~M., Tytler D., Wolfe A.~M., Kirkman D., 
Lubin D., Suzuki N., 2002b, ApJ,  571, 693 

\bibitem[\protect\citeauthoryear{Prochaska \& Wolfe}{1997}]{prochaska97} 
Prochaska J.~X., Wolfe A.~M., 1997, ApJ, 487, 73 

\bibitem[\protect\citeauthoryear{Prochaska \& Wolfe}{2002}]{prochaska02} 
Prochaska J.~X., Wolfe A.~M., 2002, ApJ, 566, 68 

\bibitem[\protect\citeauthoryear{Rauch et al.}{2008}]{rauch08} 
Rauch M., et al., 2008, ApJ, in press [arXiv:0711.1354]

\bibitem[\protect\citeauthoryear{Savage \& Lehner}{2006}]{savage06} 
Savage B.~D., Lehner N., 2006, ApJS, 162, 134 

\bibitem[\protect\citeauthoryear{Savage \& Sembach}{1991}]{savage91}
Savage, B. D.,  Sembach, K. R. 1991, ApJ, 379, 245

\bibitem[\protect\citeauthoryear{Savage \& Sembach}{1996}]{savage96} 
Savage B.~D., Sembach K.~R., 1996, ARA\&A, 34, 279

\bibitem[\protect\citeauthoryear{Savage, Sembach, \& Howk}{2001}]{savage01} 
Savage B.~D., Sembach K.~R., Howk J.~C., 2001, ApJ, 547, 907 

\bibitem[\protect\citeauthoryear{Savage et al.}{2003}]{savage03} 
Savage B.~D., et al., 2003, ApJS, 146, 125 

\bibitem[\protect\citeauthoryear{Sembach et al.}{1999}]{sembach99} 
Sembach K.~R., Savage B.~D., Lu L., Murphy E.~M., 1999, ApJ, 515, 108 

\bibitem[\protect\citeauthoryear{Sembach et al.}{2003}]{sembach03} 
Sembach K.~R., et al., 2003, ApJS, 146, 165

\bibitem[\protect\citeauthoryear{Shapley et al.}{2003}]{shapley03} 
Shapley A.~E., Steidel C.~C., Pettini M., Adelberger K.~L., 2003, ApJ, 588, 65 

\bibitem[\protect\citeauthoryear{Simcoe, Sargent, \& Rauch}{2004}]{simcoe04} 
Simcoe R.~A., Sargent W.~L.~W., Rauch M., 2004, ApJ, 606, 92 

\bibitem[\protect\citeauthoryear{Simcoe et al.}{2006}]{simcoe06} 
Simcoe R.~A., Sargent W.~L.~W., Rauch M., Becker G., 2006, ApJ, 637, 648 

\bibitem[\protect\citeauthoryear{Slavin, Shull, \& Begelman}{1993}]{slavin93} 
Slavin J.~D., Shull J.~M., Begelman M.~C., 1993, ApJ, 407, 83 

\bibitem[\protect\citeauthoryear{Sofia \& Jenkins}{1998}]{sofia98} 
Sofia U.~J., Jenkins E.~B., 1998, ApJ, 499, 951 

\bibitem[\protect\citeauthoryear{Spitzer}{1996}]{spitzer96} 
Spitzer L.~J., 1996, ApJ, 458, L29 

\bibitem[\protect\citeauthoryear{Sutherland \& Dopita}{1993}]{sutherland93}
Sutherland, R. S., Dopita, M. A. 1993, ApJS, 88, 253

\bibitem[\protect\citeauthoryear{Tremonti, Moustakas, \& Diamond-Stanic}{2007}]{tremonti07} 
Tremonti C.~A., Moustakas J., Diamond-Stanic A.~M., 2007, ApJ, 663, L77 

\bibitem[\protect\citeauthoryear{Tripp et al.}{2008}]{tripp08} 
Tripp T.~M., Sembach K.~R., Bowen D.~V., Savage B.~D., Jenkins E.~B., 
Lehner N., Richter P., 2007, ApJS, in press, [arXiv:0706.1214]

\bibitem[\protect\citeauthoryear{Tripp et al.}{2003}]{tripp03} 
Tripp T.~M., et al., 2003, AJ, 125, 3122 

\bibitem[\protect\citeauthoryear{Vladilo et al.}{2001}]{vladilo01} 
Vladilo G., Centuri{\'o}n M., Bonifacio P., Howk J.~C., 2001, ApJ, 557, 1007 

\bibitem[\protect\citeauthoryear{Vladilo et al.}{2003}]{vladilo03} 
Vladilo G., Centuri{\'o}n M., D'Odorico V., P{\'e}roux C., 2003, A\&A, 402, 487 

\bibitem[\protect\citeauthoryear{Weiner et al.}{2008}]{weiner08} 
Weiner B.~J., Coil A.~L., Newman, J.~A., Cooper M.~C., Prochaska, J.~X., Rubin, K.~H.~R., 
2008,  4th UC Irvine Center for Cosmology Workshop, submitted. 

\bibitem[\protect\citeauthoryear{White \& Frenk}{1991}]{white91} 
White S.~D.~M., Frenk C.~S., 1991, ApJ, 379, 52 

\bibitem[\protect\citeauthoryear{Wolfe \& Chen}{2006}]{wolfe06} 
Wolfe A.~M., Chen H.-W., 2006, ApJ, 652, 981 

\bibitem[\protect\citeauthoryear{Wolfe, Gawiser, \& Prochaska}{2003}]{wolfe03b} 
Wolfe A.~M., Gawiser E., Prochaska J.~X., 2003, ApJ, 593, 235 

\bibitem[\protect\citeauthoryear{Wolfe, Gawiser, \& Prochaska}{2005}]{wolfe05} 
Wolfe A.~M., Gawiser E., Prochaska J.~X., 2005, ARA\&A, 43, 861 

\bibitem[\protect\citeauthoryear{Wolfe et al.}{2004}]{wolfe04} 
Wolfe A.~M., Howk J.~C., Gawiser E., Prochaska J.~X., Lopez S., 2004, ApJ, 615, 625 

\bibitem[\protect\citeauthoryear{Wolfe \& Prochaska}{2000a}]{wolfe00a} 
Wolfe A.~M., Prochaska J.~X., 2000a, ApJ, 545, 591 

\bibitem[\protect\citeauthoryear{Wolfe \& Prochaska}{2000b}]{wolfe00b} 
Wolfe A.~M., Prochaska J.~X., 2000b, ApJ, 545, 603 

\bibitem[\protect\citeauthoryear{Wolfe, Prochaska, \& Gawiser}{2003}]{wolfe03} 
Wolfe A.~M., Prochaska J.~X., Gawiser E., 2003, ApJ, 593, 215 

\bibitem[\protect\citeauthoryear{Wolfe et al.}{2008}]{wolfe08} 
Wolfe A.~M., Prochaska J.~X., Jorgenson R.~A., Rafelski M., 2008, ApJ, in press (arXiv:0802.3914)

\bibitem[\protect\citeauthoryear{Wolfire et al.}{1995}]{wolfire95} 
Wolfire M.~G., McKee C.~F., Hollenbach D., Tielens A.~G.~G.~M., 1995, ApJ, 453, 
673 

\bibitem[\protect\citeauthoryear{Wolfire et al.}{2003}]{wolfire03} 
Wolfire M.~G., McKee C.~F., Hollenbach D., Tielens A.~G.~G.~M., 2003, ApJ, 587, 
278 

\bibitem[\protect\citeauthoryear{Yao \& Wang}{2007}]{yao07} 
Yao Y., Wang Q.~D., 2007, ApJ, 666, 242 

\bibitem[\protect\citeauthoryear{Zech et al.}{2008}]{zech07} 
Zech W.~F., Lehner, N. Howk J.~C., Dixon, W.~V.~D., Brown, T.~M., 2008, ApJ, 679, 460

\bibitem[\protect\citeauthoryear{Zsarg{\'o} et al.}{2003}]{zsargo03} 
Zsarg{\'o} J., Sembach K.~R., Howk J.~C., Savage B.~D., 2003, ApJ, 586, 1019 


\end{thebibliography}
\end{document}